\documentclass[aps,groupedaddress,showpacs,floatfix]{revtex4}
\usepackage{amssymb}
\usepackage{amsmath}
\usepackage{graphicx}
\DeclareMathOperator{\sgn}{sgn}
\DeclareMathOperator{\Ai}{Ai}
\begin{document}

\title{Universal Randomness}

\author{Victor Dotsenko$^{\, a,b}$ }

\affiliation{$^a$LPTMC, Universit\'e Paris VI, 75252 Paris, France}

\affiliation{$^b$L.D.\ Landau Institute for Theoretical Physics,
   119334 Moscow, Russia}

\date{\today}

\begin{abstract}
During last two decades it has been discovered that 
the statistical properties of a number of microscopically rather different random systems 
at the macroscopic level are described 
by {\it the same} universal probability distribution function which is
called the Tracy-Widom (TW) distribution. Among these systems we find both purely methematical
problems, such as the longest increasing subsequences in random permutations, and 
quite physical ones, such as directed polymers in random media or polynuclear crystal
growth. In the extensive Introduction we discuss in simple terms these various random systems 
and explain what the universal TW function is.
Next, concentrating on the example of one-dimensional directed polymers in random potential 
we give the main lines of the formal proof that fluctuations of their free energy
are described the universal TW distribution.
The second part of the review  consist of  
detailed appendices which provide necessary self-contained mathematical background 
for the first part.

\end{abstract}


\maketitle

\medskip

\begin{center}
 
{\bf \large Contents}

\end{center}

\begin{tabbing}
 AAAAAA \= {\bf Appendix C: Wave functions of quantum bosons with attractive interactions} \=
 . . . . \=  13\kill

\> {\bf I. Introduction}                                                 \>     \>   \\
\> \hspace{5mm}    A. Combinatorics                            \>  \>   \\
\> \hspace{5mm}    B. Polynuclear crystal growth               \>  \>   \\  
\> \hspace{5mm}    C. Directed polymers                        \>  \>   \\
\> \hspace{5mm}    D. Replica method                           \>  \>   \\
\> \hspace{5mm}    E. Tracy-Widom distribution function        \>  \>   \\
\> {\bf II. Directed polymers and one-dimensional quantum bosons}   \>   \>    \\
\> {\bf III. Solution of the one-dimensional directed polymers problem}    \>   \>    \\
\> {\bf IV. Conclusions}                                                 \>   \>    \\
\\
\> {\bf Appendix A: Quantum bosons with repulsive interactions}                     \>  \>  \\
\> \hspace{5mm}    1. Eigenfunctions                                            \>  \>  \\
\> \hspace{5mm}    2. Othonormality                                             \>  \>  \\
\> {\bf Appendix B: Quantum bosons with attractive interactions}                    \>  \>  \\
\> \hspace{5mm}    1. Ground state                                              \>  \>   \\
\> \hspace{5mm}    2. Eigenfunctions                                            \>  \>   \\
\> \hspace{5mm}    3. Othonormality                                             \>  \>   \\
\> \hspace{5mm}    4. Propagator                                                \>  \>   \\
\> {\bf Appendix C: The Airy function integral relations }                      \>  \>   \\    
\> {\bf Appendix D: Fredholm determinant with the Airy kernel}                  \>         \>     \\
\>  \hspace{23mm}  {\bf  and the Tracy-Widom distribution}                      \>  \>   \\
\end{tabbing}

%\medskip

%\vspace{5mm}

\newpage

\section{Introduction}

\newcounter{1}
\setcounter{equation}{0}
\renewcommand{\theequation}{1.\arabic{equation}}

Everyone knows the Gaussian distribution function. Whenever we are dealing with a 
system containing {\it independent} random parameters, its macroscopic characteristics
(according to the central limit theorem) are described by the Gaussian distribution.
This kind of universal behavior is trivial, and not so much interesting.
On the other hand, every non-trivial system  usually
requires individual consideration, and although there are lot of
universal macroscopic properties among microscopically different systems 
(e.g. scaling and critical phenomena at the phase transitions) 
until very recently no one would expect to have a  
{\it universal function} (different from the Gaussian one) which would describe 
macroscopic statistical properties of a whole class of non-trivial random systems. 

Originally the solution of Tracy and Widom \cite{Tracy-Widom} were devoted to 
rather specific mathematical problem, namely the distribution function of the largest eigenvalue 
of $N\times N$ Hermitian matrices (Gaussian Unitary Ensemble (GUE)) in the limit $N \to \infty$. 
Nowadays we have got rather comprehensive list of various systems (both purely mathematical 
and physical)
whose macroscopic statistical properties are described by the same universal 
Tracy-Widom (TW) distribution function.
These systems are: 
the longest increasing subsequences (LIS) model \cite{LIS} (Section I.A)
zero-temperature lattice directed polymers with geometric disorder \cite{DP_johansson}
the polynuclear growth (PNG) system \cite{PNG_Spohn}, (Section I.B)
 the oriented digital boiling model \cite{oriented_boiling}, 
the ballistic decomposition model \cite{ballistic_decomposition}, 
the longest common subsequences (LCS) \cite{LCS},
the one-point distribution of the solutions of the KPZ equation \cite{KPZ} 
(which describes the motion of 
an interface separating two homogeneous bulk phases) in the long time limit \cite{KPZ-TW1,KPZ-TW2},
and finally finite temperature directed polymers in random potentials with short-range correlations
\cite{Dotsenko1,Dotsenko2,Dotsenko3,LeDoussal}. It should be noted that
directed polymers in a quenched random potential have 
been the subject of intense investigations during the past three
decades (see e.g. \cite{hh_zhang_95}). 
Diverse physical systems such as domain walls in magnetic films
\cite{lemerle_98}, vortices in superconductors \cite{blatter_94}, wetting
fronts on planar systems \cite{wilkinson_83}, or Burgers turbulence
\cite{burgers_74} can be mapped to this model, which exhibits numerous
non-trivial features deriving from the interplay between elasticity and
disorder.

The rest of this Introduction is devoted to the discussion 
of several  random statistical systems with explanations 
in very simple terms of what the TW distribution
describes in them. Namely, we will consider the combinatorial model 
of the longest increasing subsequences (section I.A), 
the polynuclear crystal growth model (section I.B), and 
one-dimensional directed polymers in random potential (section I.C).
Besides in section I.D the main ides of the replica method (used in the present approach)
will be described, and finally in section I.E the definition and the main properties of the 
TW distribution function will be given.

Sections II and III are devoted to the exact solution of the one-dimensional
directed polymers problem. In particular, in section II the main ideas of this 
solution as well as its methodological tools are described. Section III contains 
the main lines of the derivation of the TW distribution function for the 
free energy fluctuations in one-dimensional directed polymers with $\delta$-correlated
random potential. 
The second part of this review contains several technical appendices containing 
all necessary mathematical tools (used in the previous sections) which hopefully
makes the whole paper to be self-contained.

\subsection{Combinatorics}

We start with purely mathematical "toy" model which, as we will see later, is closely related with
physical problems of polynuclear crystal growth (section I.B) and one-dimensional 
directed polymers (section I.C).
This combinatorial problem of statistical properties of the longest increasing subsequences
(LIS) was formulated long time ago by Ulam \cite{Ulam}, and hence it is often
called the Ulam's problem. Let us consider a sequence of $N$ integers 
$\{1, 2, ..., N\}$. Then, for an arbitrary permutation of these integers we have to find
all possible {\it increasing} subsequences, and among them the length $l_{N}$
of the longest ones should be defined. As an example let us consider the case $N=11$ 
and take a particular permutation
\begin{equation}
\label{1.1}
\{3, 5, 10, 1, 9, 6, 8, 4, 7, 11, 2\}.
\end{equation}
 This permutation exhibits many different  increasing 
subsequences (such as $\{3, 5, 10, 11\}$, $\{ 1, 9, 11\}$ etc.), and among them the longest ones
are $\{3, 5, 6, 7, 11\}$ and $\{3, 5, 6, 8, 11\}$. In other words, for this particular 
permutation $l_{N} = 5$. Simple graphical representation of this permutation
problem is shown in Figure 1. Here the set of $11$ bold dots inside the $(12\times 12)$ square 
represents the permutation, eq.(\ref{1.1}): for every integer in the $x$ direction one 
associates one and only one  {\it permuted} integer in the $y$ direction (for $x=1$ one gets $y=3$, 
for $x=2$ one gets $y=5$, etc.). All possible increasing subsequences of this particular 
permutation are obtained by drawing all possible {\it directed paths} connecting the 
origin $(0,0)$ with the right up corner $(12,12)$ of this square, which are passing over 
the internal bold dots. Directed path means that only ``right-and-up`` movements are allowed when 
going from one dot to another. For example, from the point $(6,6)$ one can jump only to the 
points $(7,8)$, $(9,7)$ and $(10,11)$. In other words, when going from the origin to the 
right up corner $(12,12)$ both $x$ and $y$ coordinates can only increase at every step.
In terms of these rule, the longest increasing subsequence is described by the directed
path which goes over the maximum possible number of dots. Note that for a given permutation 
the longest increasing subsequence is not necessary unique. In the considered example
in addition to the subsequence  $\{3, 5, 6, 7, 11\}$ shown in Figure 1 by the dotted line,
there exist another one namely $\{3, 5, 6, 8, 11\}$.
\begin{figure}[h]
\begin{center}
   \includegraphics[width=8.0cm]{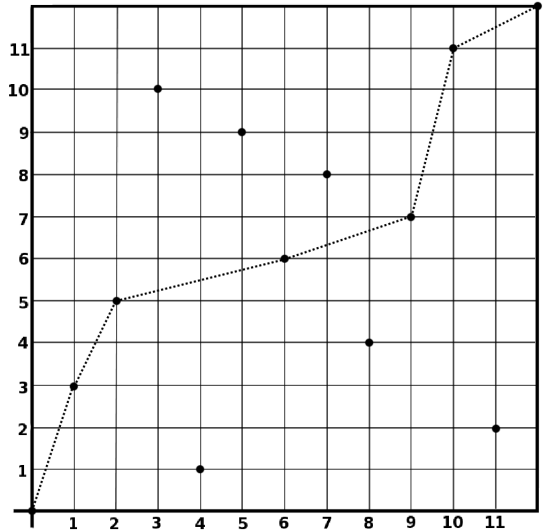}
\caption[]{Geometrical representation of the permutation, eq.(\ref{1.1}), for $N=11$. 
The dotted line corresponds the longest increasing subsequence $\{3, 5, 6, 7, 11\}$ }
\end{center}
\label{figure1}
\end{figure}

Considering all $N!$ possible permutations with equal probability one find that 
the length $l_{N}$ varies from permutation to permutation being the random variable.
The question is what are the statistical properties of the random quantity $l_{N}$.
First it has been shown that in the limit of large $N$, the average value
of the longest increasing subsequence, $\overline{l_{N}}$, is proportional to
$\sqrt{N}$, namely $\overline{l_{N}} \sim c \sqrt{N}$, where
the constant $c=2$ \cite{Vershik-Kerov}. Moreover, in the limit $N\to\infty$, 
the quantity $l_{N}$ in order $\sqrt{N}$ is {\it selfaveraging}: 
$\lim_{N\to\infty} l_{N}/\sqrt{N} \, = \, 2$ (in other words, the distribution function
of the ratio $l_{N}/\sqrt{N} $ in the limit  $N\to\infty$ shrinks to the 
$\delta$-function). 

Note that at the qualitative level one can easily understand why the typical value
of $l_{N}$ must be proportional to $\sqrt{N}$. Indeed, for large $N$ a generic permutation
of the numbers $\{1, 2, ..., N\}$ in terms of the permutation matrix of Figure 1
will be represented by a {\it uniform} distribution of $N$ dots inside the $N\times N$ square.
 Thus the density of the dots will be proportional to $1/N$ while the typical distance
between neighboring dots must be proportional to $\sqrt{N}$. It means that the typical
number of dots on a diagonal-like path of the length $\sim N$ must be proportional to
$N/\sqrt{N} = \sqrt{N}$.

However, the main interest in this system is not the typical value of the
longest increasing subsequences, but their {\it fluctuations}. Recently it has been shown 
\cite{LIS,Aldous} that in the limit of large $N$ the fluctuations of $l_{N}$ scale as $N^{1/6}$,
namely,
\begin{equation}
\label{1.2}
l_{N} \; \simeq \; 2\sqrt{N} \, + \, N^{1/6} \, s
\end{equation}
where the random quantity $s$ is described by the universal $N$-independent 
distribution function $P_{TW}(s)$, which is the Tracy-Widom distribution
(see section I.E): 
\begin{equation}
\label{1.3}
\lim_{N\to\infty} \mbox{Prob} \biggl(\frac{l_{N} - 2\sqrt{N}}{N^{1/6}} = s\biggr) \; = \; 
P_{TW}(s)
\end{equation}

It turns out that the above purely mathematical "toy" model is equivalent
to the physical model of (2+1)-dimensional polynuclear crystal growth, where the 
TW distribution describes the fluctuations of the number of the crystal mono-layers 
(see next subsection).

\subsection{Polynuclear crystal growth}

It turns out that the mathematical "toy" model considered above is equivalent to the 
physical model of the crystal growth with randomly located nucleation centers.
This is the model of {\it polynuclear crystal growth} (PNG) which describes the 
growth of the two-dimensional crystal monolayers in (2+1) dimensions.

Let us consider again Figure 1 where the bold dots will be assumed to
represent the nucleation centers. The crystal layers growth takes place
in the vertical direction (toward the reader) according to the following rules.
From each nucleation center we draw the monolayer level step straight line in the horizontal
direction to the right and in the vertical direction up, until these lines meet with 
the other lines starting from the other centers. In this way we are getting the monolayer
"terraces" which mount from the left-down to the right-up corner of the square.
(see Figure 2). 
\begin{figure}[h]
\begin{center}
   \includegraphics[width=10.0cm]{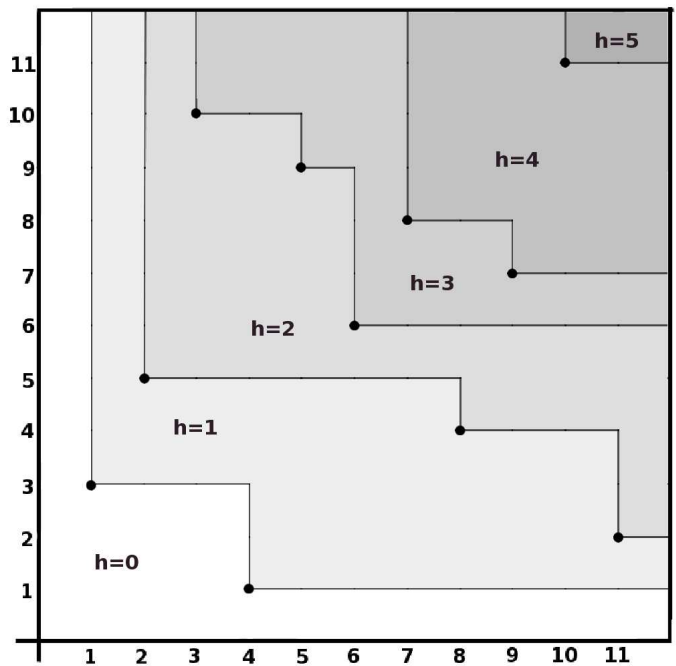}
\caption[]{(2+1)-dimensional crystal terraces with the "nucleation centers"
corresponding to the permutation model of Figure 1}
\end{center}
\label{figure2}
\end{figure}
One can easily see that for a given random positions of the nucleation centers
(for a given permutation in the previous LIS problem) the number of terraces
$h_{N}$ is just equal to the longest increasing permutation $l_{N}$ in the 
combinatorial problem considered in the previous subsection. For a given value of $N$,
depending on the actual configurations of the nucleation centers inside the 
$N\times N$ square, the number of the monolayer terraces is the random
quantity, and in the limit $N\to\infty$ its statistics is described by
the TW distribution, eq.(\ref{1.3}) \cite{PNG_Spohn}. 
At present the study of various modifications of PNG model formulated above
is the vast field of research (see e.g. \cite{Ferrari}). It is also interesting to
note that quite recently the existence of TW distribution in the PNG-like 
systems has been confirmed experimentally \cite{takeuchi}.

\subsection{Directed polymers}

It is clear that when the size of the square $N$ is large the presence
of the background lattice is not essential. In this case instead of 
considering the problem in terms of permutations one can introduce a
homogeneous distribution of dots inside continuous square (Figure 3).
\begin{figure}[h]
\begin{center}
   \includegraphics[width=10.0cm]{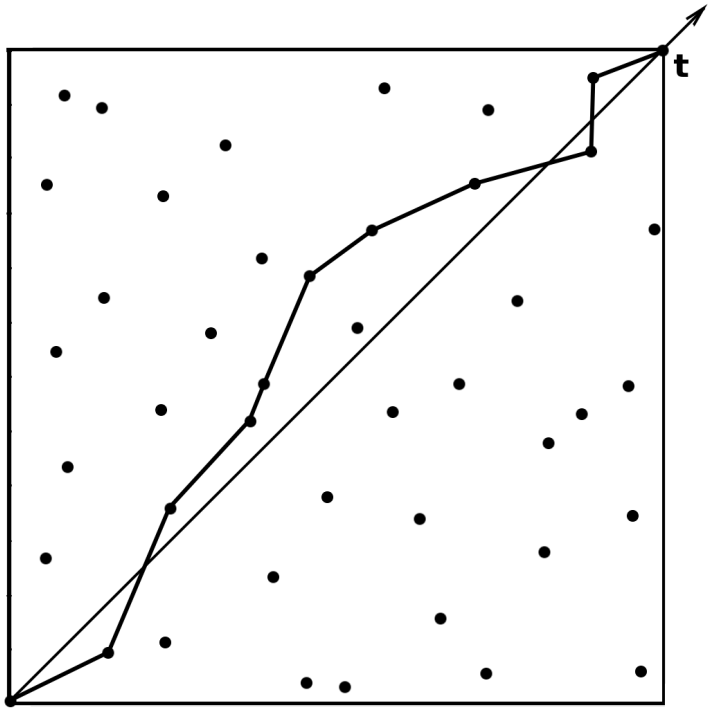}
\caption[]{Directed path through  randomly distributed dots}
\end{center}
\label{figure3}
\end{figure}
Let us introduce here the diagonal "time" axes which goes from the origin
(left down corner) to the right up corner of the square. The directed polymer
here is the line which starts at the origin and arrives to the right up corner
"jumping" over the dots in such a way that its time coordinates increases at
every jump. For a given (random) configuration of dots among many possible
directed polymer trajectories we have to choose the ones which contains
the maximum number of dots.  In this formulation
the problem of statistics of the length of such directed polymers
looks somewhat different from the LIS problem considered above. It can be shown
however, that in the limit of large times $t$ and large $N$ these two problems
become equivalent \cite{LIS}. 
For a given (fixed) density $\rho$ of dots instead of the total number of dots $N$ 
one can measure the length $l(t)$ of the polymer in terms of the size of the square $t$.
Since $\rho = 2N/t^{2}$, assuming that  $\rho$  is a parameter which is of the order
of one, we note that  $t \propto \sqrt{N}$. In this case 
we find that the fluctuations of the length of the polymers scales as $t^{1/3}$, and
instead of eq.(\ref{1.3}) we get
\begin{equation}
\label{1.4}
\lim_{t\to\infty} \mbox{Prob} \biggl(\frac{l(t) - \sqrt{2\rho} \, t}{t^{1/3}} = s \biggr) \; = \; 
P_{TW}(s)
\end{equation}
In other words in the {\it thermodynamic limit}, $t\to\infty$ all the systems considered above
appear  to be equivalent to each other which doesn't look so much surprising if we compare
Figures 1-3. 

In statistical physics one defines random directed polymers
in somewhat different way. Let us introduce a square lattice in which discrete "time"
$t = 1, 2, ..., L$ is now horizontal (Figure 4). The vertical direction is described by 
the discrete parameter $\phi =0, \pm 1, \pm 2, ..., \pm M$. At every lattice site $(\phi, t)$
we place random quantities (random potential) $V(\phi, t)$ and assume that they are described by 
{\it independent} Gaussian distributions:
\begin{equation}
\label{1.5}
{\cal P}[{\bf V}] \; = \; \prod_{\phi, t} 
\sqrt{\frac{1}{2\pi u}} \exp\Bigl(-\frac{1}{2u} V^{2}(\phi, t)\Bigr)
\end{equation}
The parameter $u$ defines the typical strength of the random potentials $V(\phi, t)$, which
according to eq.(\ref{1.5}) are uncorrelated
\begin{equation}
\label{1.6}
\overline{V(\phi, t) \, V(\phi', t')} \; = \; u \, \delta_{\phi, \phi'} \, \delta_{t, t'}
\end{equation}
and have zero mean value, $\overline{V(\phi, t)} = 0$ (the horizontal line denotes the averaging
with the distribution, eq.(\ref{1.5})).
\begin{figure}[h]
\begin{center}
   \includegraphics[width=10.0cm]{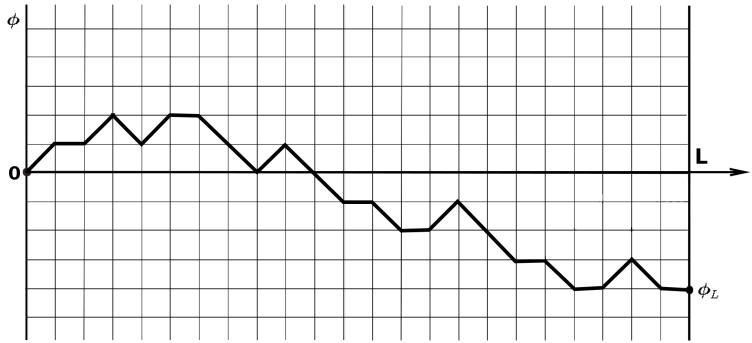}
\caption[]{Directed polymer on a square lattice}
\end{center}
\label{figure4}
\end{figure}

The directed polymer here is the path which starts at the origin and goes over the lattice sites 
to the right end of the system. At every time step, $t \to t+1$ the polymer trajectory $\phi(t)$, 
can deviate up or down by one step or may not deviate at all: 
$\phi(t+1) = \phi(t) + \sigma(t+1)$, where $\sigma = \pm 1, 0$. Assuming that the polymer is a kind of
elastic string  we can introduce  
 "elastic" (positive) energy $\propto \bigl[\phi(t+1) - \phi(t)\bigr]^{2}$ for every polymer's
deviation. In this way for a given trajectory $\phi(t)$   we can associate the 
following {\it energy}:
\begin{equation}
\label{1.7}
H[\phi(t)]\; = \; \sum_{t=1}^{L-1} \Bigl[\frac{1}{2} \bigl[\phi(t+1) - \phi(t)\bigr]^{2}
+ V(\phi(t), t) \Bigr]
\end{equation}
This expression contains two competing contributions: the first (elastic) terms are trying to make
the trajectory as horizontal as possible, while the second ones are forcing the trajectory
to deviate in the search for the most negative values of the random potentials. For a given
(random) configuration of the potentials $V(\phi, t)$ the optimal trajectory $\phi_{*}(t)$ 
is defined by
the minimum of the total energy, eq.(\ref{1.7}), 
\begin{equation}
\label{1.8}
E(L)\; = \; \min_{\phi(t)} \Biggl\{\sum_{t=1}^{L-1} \Bigl[\frac{1}{2} \bigl[\phi(t+1) - \phi(t)\bigr]^{2}
+ V(\phi(t), t) \Bigr] \Biggr\}
\end{equation}
Being the function of the random potential $V(\phi, t)$ this quantity is also random, and 
it could be considered as the distant analog of the (random) length of the directed polymers in the previous
example shown in Figure 3. An essential difference is that in this latter case the elastic terms
are absent, while the (negative) contribution of the potential energy is associated with
a fixed energy $V_{0}$ carried by the dots which (unlike the Gaussian random potentials $V(\phi, t)$)
are {\it geometrically} random. One more important difference is that unlike the directed polymer
in Figure 4, which is defined by the {\it local} in space one-step wandering, the trajectory
in Figure 3 can jump {\it any distance} from dot to dot (not necessary between neighboring dots).
Thus, {\it a priori} there are no reasons to expect that these two quantities, $E(L)$, eq.(\ref{1.8}),
and $l(t)$ in the example of Figure 3 would have the same statistical properties. 
If, nevertheless, we would {\it suppose} that at least in the thermodynamic limit, $t \sim L \to \infty$, 
these two types of systems become equivalent we would have to expect that 
$E(L) \; \simeq \; f_{0}\, L \; + \; L^{1/3} s$, where $f_{0}$  is the {\it bulk} 
(selfaveraging) 
energy density and  the random parameter $s$ is described by $L$-independent
TW distribution function. 

The system defined by the Hamiltonian (\ref{1.7}) 
is the usual one-dimensional statistical system
containing quenched disorder. 
The fact that the leading contribution to its 
ground state energy $E(L)$ is proportional to the system size $L$ 
can be explained in very simple way. Indeed, 
in the first approximation, to minimize the energy at every time step
among three possible options ("up","horizontal" and "down") 
the trajectory of Figure 4 can choose the site
where the value of the random potential $V(\phi, t)$ is lower 
(in this approximation we neglect the presence of 
non-local phenomena when the optimal trajectory chooses 
locally unfavorable option to gain globally more favorable
energy). In this way the second term in eq.(\ref{1.7}) 
provides the contribution which is
proportional to $-\sqrt{u}\, L$ (and not $\sqrt{u L}$,
 as it would be for an {\it arbitrary} trajectory $\phi(t)$).
Since the contribution of the first (elastic) term in eq.(\ref{1.7}) 
is also proportional to $L$, we find that 
in the leading order in $L$, the energy of the optimal trajectory 
$E(L) \; \simeq \; -(const)\, L $, and moreover,
we can be sure that this contribution is negative since 
the energy of the optimal trajectory in the absence
on the random potentials is zero (it is just the straight horizontal line), 
while the presence of the random
potential can only lower the energy. 
On the other hand the fact that {\it finite-size corrections}  
in this system are of  order $L^{1/3}$ 
(and {\it not} of order $L^{1/2}$, as one could naively expect) 
is highly non-trivial phenomenon 
which is very difficult to explain in simple terms.

The lattice model as it is introduced above, eqs.(\ref{1.7})-(\ref{1.8}), 
is essentially the zero-temperature system, as we are dealing here with the optimal 
(global minimum) trajectories only.
It is clear that the search for the global minimum configurations in eq.(\ref{1.8}) is highly 
non-trivial task, as it can not be done via the local in time algorithms. On the other hand, 
as it often happens, one can make life much easier if one consider more general (i.e. more 
complicated) problem. Namely, let us introduce finite {\it temperature} $T$ in the 
system so that in addition to the quenched disorder fluctuations we would have 
the contributions of the thermal fluctuations produced by the trajectories away from 
the global minima ones. In terms of this generalization instead of the global 
minimum energy $E(L)$, 
eq.(\ref{1.8}), we would get the {\it free energy}:
\begin{equation}
\label{1.9}
F(L, T)\, = \, - T \ln\Bigl[\sum_{\phi(t)} \exp\Bigl(-\frac{1}{T} H[\phi(t)]\Bigr) \Bigr]
\end{equation} 
where the expression under the logarithm is the partition function in which the summation goes over
{\it all} trajectories starting at the origin. In the case the global minimum trajectory is unique 
(which is usually the case in the large system) the energy defined in eq.(\ref{1.8}) is obtained
by taking the zero temperature limit: $E(L) \, = \, \lim_{T\to 0} F(L,T)$.
The finite temperature lattice model defined by eqs.(\ref{1.7}),(\ref{1.9}) is sufficiently
simple for numerical investigations (see e.g.  \cite{Krug}) which in particular allows to demonstrate
the existence of the free energy fluctuations scaling $\sim L^{1/3}$.  On the other hand, the problem
defined on a lattice is very hard for analytical studies. 
Since the phenomena of interest are taking place at
large system sizes, one may hope that it would be  sufficient to consider the system in the 
continuous limit where the presence of a lattice become irrelevant. Continuous limit generalization 
of the Hamiltonian, eq.(\ref{1.7}) is straightforward. Assuming that the lattice spacing goes to zero 
and changing the finite differences in its first
term by the gradients we get
\begin{equation}
\label{1.10}
H[\phi(t)]\; = \; \int_{0}^{L} d\tau \biggl\{\frac{1}{2} \biggl[\frac{d\phi(\tau)}{d\tau}\biggr]^{2}
+ V\bigl(\phi(\tau), \tau\bigr) \biggr\}
\end{equation}
where, as before, the disorder potential $V(\phi, \tau)$ is Gaussian distributed and 
uncorrelated. Instead of eq.(\ref{1.6}) in the continuous limit we get
\begin{equation}
\label{1.11}
\overline{V(\phi, \tau) \, V(\phi', \tau')} \; = \; u \, \delta(\phi - \phi') \, \delta(\tau-\tau')
\end{equation}
The partition function of this system is now defined in terms of the functional
integral:
\begin{equation}
\label{1.12}
   Z = \int_{-\infty}^{+\infty} d\phi_{L}\int_{\phi(0)=0}^{\phi(L)=\phi_{L}} 
              {\cal D} [\phi(\tau)]  \;  \mbox{\Large e}^{-\beta H[\phi]}
\end{equation}
where $\beta = 1/T$ is the inverse temperature and the integration goes over all trajectories 
starting at the origin ($\tau=0$) and having free boundary conditions at $\tau = L$.
In this way instead of the lattice trajectory of Figure 4 we are getting 
a continuous trajectory
shown in Figure 5.
Although, as we will see later, the continuous model defined above is ill defined at short
distances, as far as its long-time behavior is concerned it is much 
better treatable  analytically.
\begin{figure}[h]
\begin{center}
   \includegraphics[width=10.0cm]{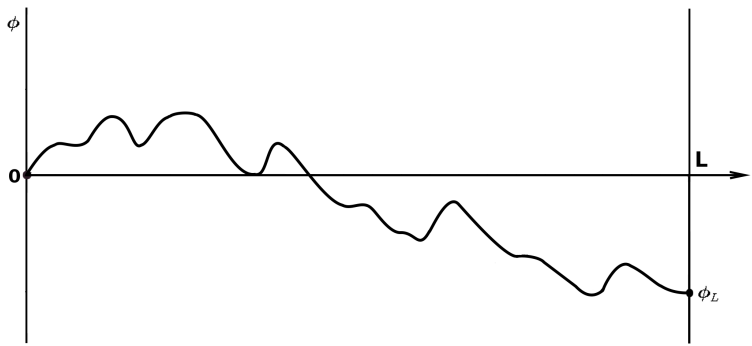}
\caption[]{Continuous elastic string in a random potential, eq.(\ref{1.10})}
\end{center}
\label{figure5}
\end{figure}

First of all we can note that in the absence of the random potential in the Hamiltonian,
eq.(\ref{1.10}), the system describes simple thermal diffusion. Indeed, the probability
that at time $\tau = L$ the trajectory arrives to the point $\phi(L) = \phi_{L}$ is given by
the partition function
\begin{equation}
\label{1.13}
   Z_{0}(\phi_{L}) = \int_{\phi(0)=0}^{\phi(L)=\phi_{L}} 
              {\cal D} [\phi(\tau)]  \; 
\exp\biggl\{-\frac{1}{2T} \int_{0}^{L} d\tau \biggl[\frac{d\phi(\tau)}{d\tau}\biggr]^{2} \biggr\}
\end{equation}
Simple Gaussian integration (with the proper choice of the {\it integration measure} 
of the functional integral) yields
\begin{equation}
\label{1.14}
   Z_{0}(\phi_{L}) = \frac{1}{\sqrt{2\pi T L}} \exp\biggl\{-\frac{\phi_{L}^{2}}{2T L} \biggr\} 
\end{equation}
In other words the typical deviation $<\phi_{L}>$ of the trajectory due to the {\it thermal
fluctuations} growth as $\sqrt{T} \,  L^{1/2}$ (and it shrinks to zero at $T=0$). 
On the other hand, in the presence of the random potential  
the two terms of the Hamiltonian (\ref{1.10}) must balance each other.
For a given value of the typical deviation $\phi_{L}$ the contribution of the elastic term 
can be estimated  as $\phi_{L}^{2}/ L$. Thus, if in the presence of disorder the free energy 
fluctuations of this system are scaling as $L^{1/3}$ (see e.g. \cite{hhf_85,numer1,numer2,kardar_87}) 
we can conclude that the typical value of 
the trajectory deviations due to the action of the random potentials must grow 
as $\phi_{L} \sim L^{2/3}$ 
which is much faster than the pure thermal diffusion scaling $L^{1/2}$. 

In what follows we are going to study the system defined by eqs.(\ref{1.10}),(\ref{1.11})
in all detail. It turns out that regardless of essential differences in the 
definition  of this system and the previous models discussed above in sections 1.A-1.C,
in the thermodynamic limit all these models becomes {\it equivalent}.
The central result which will be proved in the next sections is in the following.
In the limit $L\to\infty$ the free energy of the considered system can be
represented as
\begin{equation}
\label{1.15}
F \; = \; f_{0} L \; + \; c \, L^{1/3}\; f
\end{equation}
where $f_{0}$ is the linear free energy density, the constant
$c = \frac{1}{2}(\beta^{2}u^{2})^{1/3}$, and  the random quantity
$f \sim 1$ (just like the quantity $s$ in eqs.(\ref{1.2})-(\ref{1.4})) is described 
by the universal TW distribution function (see section I.E)

\subsection{Replica method}

The replica method is widely used in the studies of systems containing quenched disorder
(see e.g. \cite{ufn,book}).
For simplicity let us consider the  string $\phi(\tau)$ with the zero boundary conditions:
$\phi(0) = \phi(L) = 0$. The partition function of a given sample described by the 
Hamiltonian, eq.(\ref{1.10}), is
\begin{equation}
\label{1.16}
   Z[V] = \int_{\phi(0)=0}^{\phi(L)=0} 
              {\cal D} [\phi(\tau)]  \;  \mbox{\Large e}^{-\beta H[\phi,V]}
\end{equation}
On the other hand,
the partition function is related to the total free energy $F[V]$ via
\begin{equation}
\label{1.17}
Z[V] = \exp( -\beta  F[V])
\end{equation}
The free energy $F[V]$  is defined for a specific 
realization of the random potential $V$ and thus represent a random variable. 
Taking the $N$-th power of both sides of this relation 
and performing the averaging over the random potential $V$ we obtain
\begin{equation}
\label{1.18}
\overline{Z^{N}[V]} \equiv Z[N,L] = \overline{\exp( -\beta N F[V]) }
\end{equation}
where the quantity in the lhs of the above equation is called the 
{\it replica partition function}.
Substituting here the free energy in the form 
$F = f_{0} L +  c \, L^{1/3}\; f$, eq.(\ref{1.15}),
 and redefining the partition function
\begin{equation}
\label{1.19}
Z[N,L] = \tilde{Z}[N,L] \; \mbox{\Large e}^{-\beta N f_{0} L} 
\end{equation}
we get
\begin{equation}
\label{1.20}
\tilde{Z}[N,L] = \overline{\exp( -\lambda N f) }
\end{equation}
where $\lambda = \beta c L^{1/3}$.
The averaging in the rhs of the above equation can be represented in terms of the 
distribution function $P_{L}(f)$ (which depends on the system size $L$). 
In this way we arrive to the following general relation 
between the replica partition function $\tilde{Z}[N,L]$ and the distribution function 
of the free energy fluctuations $P_{L}(f)$:
\begin{equation}
\label{1.21}
   \tilde{Z}[N,L] \; =\;
           \int_{-\infty}^{+\infty} df \, P_{L} (f) \;  
           \mbox{\Large e}^{ -\lambda N   \, f}
\end{equation}
Of course, the most interesting object is the thermodynamic limit 
distribution function $P_{*}(f) = \lim_{L\to\infty} P_{L} (f)$ which is expected to be 
the universal quantity. The above equation is the bilateral Laplace transform of  the function $P_{L}(f)$,
and at least formally it allows to restore this function via inverse Laplace transform 
 of the replica partition function $\tilde{Z}[N,L]$. In order to do so one has to compute 
$\tilde{Z}[N,L]$ for an  {\it arbitrary }
integer $N$ and then perform  analytical continuation of this function 
from integer to arbitrary complex values of $N$. This is the standard  strategy of the replica method
in disordered systems where it is well known that  very often the procedure of such analytic
continuation turns out to be rather controversial point  \cite{zirnbauer1,replicas}. Even in rare  cases when 
the derivation of the replica partition function $Z(N)=\overline{Z^{N}}$  can be done exactly, its
further analytic continuation to non-integer $N$ appears  to be ambiguous.
The classical example of this situation is provided by the Derrida's Random Energy Model
 in which the momenta $Z(N)$ growths as fast as  $\exp(N^2)$ at large $N$, and in this case
there are many different distributions yielding the same values of $Z(N)$, but 
providing {\it different} values for the average free energy of the system \cite{REM}.
In our present system the situation is even worse because, as we will see later, the replica
partition function growth here as  $\exp(N^3)$ at large $N$, and in this situation its
analytic continuation from integer to non integer $N$ would be rather problematic point.
It turns out, however,  that in our present case it is possible to bypass the problem 
of the analytic continuation if instead of the
distribution function $P_{*}(f)$ one would study its integral
representation
\begin{equation}
 \label{1.22}
W(x) \; = \; \int_{x}^{\infty} \; df \; P_{*}(f)  
\end{equation}
which gives the probability to find the fluctuation $f$ bigger that a given value $x$.
Formally the function $W(x)$ can be defined as follows:
\begin{eqnarray}
 \label{1.23}
W(x) &=& \lim_{\lambda\to\infty} \sum_{N=0}^{\infty} \frac{(-1)^{N}}{N!} 
\exp(\lambda N x) \; \overline{\tilde{Z}^{N}}
\\
\nonumber 
&=& \lim_{\lambda\to\infty} \sum_{N=0}^{\infty} \frac{(-1)^{N}}{N!} 
\; \overline{\exp(\lambda N x - \lambda N f)}
\\
\nonumber 
& = &  \lim_{\lambda\to\infty} \overline{\exp\bigl[-\exp\bigl(\lambda (x-f)\bigr)\bigr]} 
\; = \;  
\overline{\theta(f-x)}
\end{eqnarray}
Thus, the probability function $W(x)$ 
can be computed in terms of the above replica partition function
$\tilde{Z}[N,L]$ by summing over all replica {\it integers}
\begin{equation}
 \label{1.24}
W(x) \; = \; \lim_{\lambda\to\infty} \sum_{N=0}^{\infty} \frac{(-1)^{N}}{N!} 
\exp(\lambda N x) \; \tilde{Z}[N,L]
\end{equation}
Of course, keeping in mind that $\tilde{Z}[N,L] \sim \exp(N^3)$ at large $N$, we see that
the above series is not that innocent. Here in accordance with the {\it troubles conservation law}
instead of the problem of analytic continuation we are facing formally divergent series.
Nevertheless, it can be shown that this {\it sign alternating} series  can be
regularized in the standard way (similarly to formally divergent sign alternating series
$\sum_{k=0}^{\infty} (-1)^{k} a^{k} \; = \; (1+a)^{(-1)}$ which at $|a| > 1$ is defined as
the analytic continuation from the region $|a| < 1$). This eventually allows to
prove that the thermodynamic limit function $W(x)$, eq.(\ref{1.24}), 
is defined by the universal
Tracy-Widom distribution function.

\subsection{Tracy-Widom distribution function}

Originally the Tracy-Widom distribution function has been derived in the context 
of the statistical properties of the Gaussian Unitary Ensemble (GUE) 
of random Hermitian matrices
\cite{Tracy-Widom}. GUE is the set of $N\times N$ 
random complex Hermitian matrices $G_{ij}$
(such that $G_{ij} = G_{ji}^{*}$) whose elements are drawn independently from the Gaussian
distribution
\begin{equation}
\label{1.25}
{\cal P}[{\bf G}] \; = \; B_{N} \, \exp\Bigl\{ -\frac{1}{2} Tr({\bf G}^{2}) \Bigr\}
\end{equation}
where $B_{N}$ is the normalization constant. The joint probability density of $N$ eigenvalues
$\{\lambda_{1}, \lambda_{2}, ..., \lambda_{N}\} $
of such matrices has rather compact form \cite{Wigner}:
\begin{equation}
\label{1.26}
{\cal P}[\lambda_{1}, \lambda_{2}, ..., \lambda_{N}] \; = \; C_{N} \, 
\prod_{i\not= j}^{N} \Big|\lambda_{i} - \lambda_{j}\Big|^{2} 
\exp\Bigl\{ -\sum_{i=1}^{N} \lambda_{i}^{2} \Bigr\}
\end{equation}
where $C_{N}$ is the normalization constant. Using this joint probability 
density one can calculate various averaged characteristics of the 
eigenvalue statistics. For example, one can introduce the average
density of the eigenvalues 
$\rho(\lambda, N) \; = \; \frac{1}{N} \sum_{i=1}^{N}
 \langle \delta(\lambda - \lambda_{i})\rangle$ where the averaging 
$\langle .....\rangle$ is performed with the probability distribution,
eq.(\ref{1.26}). Using the symmetry of this distribution one gets
\begin{equation}
\label{1.27}
\rho(\lambda, N) \; = \; 
\biggl[\prod_{i=2}^{N} \int_{-\infty}^{+\infty} \; d\lambda_{i} \biggr]
{\cal P}[\lambda, \lambda_{2}, ..., \lambda_{N}] 
\end{equation}
It can be shown \cite{Wigner} that in the limit of large $N$
\begin{equation}
\label{1.28}
\rho(\lambda, N) \; = \; \sqrt{\frac{2}{N \pi^{2}} \; 
\biggl(1 - \frac{\lambda^{2}}{2 N} \biggr) }
\end{equation}
We see that on average the eigenvalues lie within the 
finite interval $\bigl[ -\sqrt{2N} < \lambda < \sqrt{2N} \bigr]$
where, according to eq.(\ref{1.28}), their density has the semi-circular form.
This is one of the central results of the random matrix theory which is 
called the Wigner semi-circular law. In particular this result tells that
on average the maximum eigenvalue $\lambda_{max}$ is equal to $\sqrt{2N}$. 
However, at large but finite $N$ the value of $\lambda_{max}$ is the random quantity
which fluctuates from sample to sample. One may ask, what is the full probability
distribution  of the largest eigenvalue  $\lambda_{max}$?
This distribution can be computed in terms of the general probability density,
eq.(\ref{1.26}). Introducing standard notations of the random matrix theory
we define the function 
$F_{2}(s) \equiv \mbox{Prob}\bigl[\lambda_{max} < s\bigr]$ which gives the probability that
$\lambda_{max}$ is less than a given value $s$ 
(in these notations the functions $F_{1}(s)$, $F_{2}(s)$ and $F_{4}(s)$ 
denote the probability distributions of the largest eigenvalues in the 
Gaussian Orthogonal Ensemble (GOE), Gaussian Unitary Ensemble (GUE) 
and Gaussian Symplectic Ensemble (GSE) correspondingly \cite{Tracy-Widom2}).
By definition
\begin{equation}
\label{1.29}
F_{2}(s) \;  = \; 
\biggl[\prod_{i=i}^{N} \int_{-\infty}^{s} \; d\lambda_{i} \biggr]
{\cal P}[\lambda_{1}, \lambda_{2}, ..., \lambda_{N}] \; \equiv \; 
 \int_{-\infty}^{s} d\lambda \, P_{TW}(\lambda) 
\end{equation}
It is this problem which has been solved by Tracy and Widom in 1994 \cite{Tracy-Widom}. 
It has been shown that at large $N$ the typical fluctuations of $\lambda_{max}$ around
its mean value $\sqrt{2N}$ scale as $N^{-1/6}$, namely (c.f. eq.(\ref{1.2}))
\begin{equation}
\label{1.30}
\lambda_{max} = \sqrt{2N} \; + \; \frac{1}{\sqrt{2} \, N^{1/6} } \, s
\end{equation}
where the random quantity $s$ is described by $N$-independent distribution
$P_{TW}(s) = dF_{2}(s)/ds$. The function $F_{2}(x)$, has the following 
explicit form\
\begin{equation}
\label{1.31}
F_{2}(s) \; = \; \exp\biggl(-\int_{s}^{\infty} dt \; (t-s) \; q^{2}(t)\biggr)
\end{equation}
or
\begin{equation}
 \label{1.32}
P_{TW}(s) \; = \; \frac{d}{ds} F_{2}(s) \; = \; 
\exp\biggl[
-\int_{s}^{\infty} dt \, (t-s) q^{2}(t) 
\biggr] \times \int_{s}^{\infty} dt \, q^{2}(t)
\end{equation}
where the function $q(t)$ is the solution of 
the Panlev\'e II equation$^{(1)}$\footnotetext[1]{There exist six Panlev\'e differential
equations which were discovered about a hundred years ago\cite{Panleve} 
(for the recent review
see e.g. \cite{Clarkson}). It is proved that the general solutions 
of the Panleve\'e equations 
are transcendental in a sense that they can not be expressed in terms 
of any of the previously 
known function including all classical special functions. At present the Panlev\'e 
equations have many applications in various parts of modern physics including
statistical mechanics, plasma physics, nonlinear waves, quantum field theory
and general relativity}, 
\begin{equation}
 \label{1.33}
q'' = t q + 2 q^{3}
\end{equation}
with the boundary condition, $q(t\to +\infty) \sim Ai(t)$. 
\begin{figure}[h]
\begin{center}
   \includegraphics[width=8.0cm]{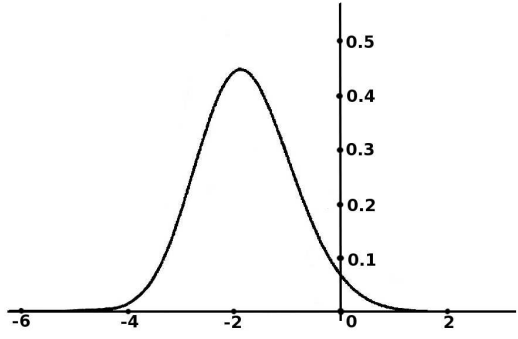}
\caption[]{Tracy-Widom distribution function $P_{TW}(x)$, eq.(\ref{1.32}).}
\end{center}
\label{figure6}
\end{figure}
The shape of the function $P_{TW}(s)$ is shown in Figure 6. Note that the asymptotic tails 
of this function are strongly asymmetric. While its right tail coincides the Airy function
asymptotic $P_{TW}(s \to +\infty) \sim \exp\bigl[-\frac{4}{3} s^{3/2}\bigr]$, the left tail
exhibits much faster decay 
$P_{TW}(s\to -\infty) \sim  \exp\bigl[-\frac{1}{12} |s|^{3}\bigr]$

\section{Mapping to quantum bosons}

\newcounter{2}
\setcounter{equation}{0}
\renewcommand{\theequation}{2.\arabic{equation}}

Explicitly, the replica partition function, Eqs.(\ref{1.18}), (\ref{1.16}), 
of the system described by
the Hamiltonian, Eq.(\ref{1.10}), is
\begin{equation}
\label{2.1}
   Z(N,L) = \prod_{a=1}^{N} \int_{\phi_{a}(0)=0}^{\phi_{a}(L)=0} 
   {\cal D} \phi_{a}(\tau) \;
   \overline{\exp\Biggl[-\beta \int_{0}^{L} d\tau \sum_{a=1}^{N}
   \bigl\{\frac{1}{2} \bigl[\partial_\tau \phi_{a}(\tau)\bigr]^2 
   + V[\phi_{a}(\tau),\tau]\bigr\}\Biggr] }
\end{equation}
Since  the random potential $V[\phi,\tau]$ has the Gaussian 
distribution the disorder averaging $\overline{(...)}$ in the above equation 
is very simple:
\begin{equation}
\label{2.2}
\overline{\exp\Biggl[-\beta \int_{0}^{L} d\tau \sum_{a=1}^{N}
     V[\phi_{a}(\tau),\tau]\Biggr] } \; = \; 
\exp\Biggl[\frac{\beta^{2}}{2} \int\int_{0}^{L} d\tau d\tau' \sum_{a,b=1}^{N}
     \overline{V[\phi_{a}(\tau),\tau] V[\phi_{b}(\tau'),\tau']}\Biggr]
\end{equation}
Using Eq.(\ref{1.11}) we find:
\begin{equation}
\label{2.3}
   Z(N,L) = \prod_{a=1}^{N} \int_{\phi_{a}(0)=0}^{\phi_{a}(L)=0} 
   {\cal D} \phi_{a}(\tau) \;
   \exp\Biggl[-\frac{1}{2}\beta \int_{0}^{L} d\tau 
   \Bigl\{\sum_{a=1}^{N} \bigl[\partial_\tau \phi_{a}(\tau)\bigr]^2 
   -\beta u \sum_{a,b=1}^{N} \delta\bigl[\phi_{a}(\tau)-\phi_{b}(\tau)\bigr]\Bigr\}\Biggr] 
\end{equation}
It should be noted that the second term in the exponential 
of the above equation contain formally divergent contributions proportional
to $\delta(0)$ (due to the terms with $a=b$). In fact, this is just an indication
that the {\it continuous} model, Eqs.(\ref{1.10})-(\ref{1.11}) is ill defined
as short distances and requires proper lattice regularization. Of course, the 
corresponding lattice model  Eqs.(\ref{1.6})-(\ref{1.7}) contains no divergences, and the terms with
$a=b$ in the exponential of the corresponding replica partition function would produce
irrelevant constant $\frac{1}{2}L\beta^2 u N \delta(0)$ (where the lattice
version of $\delta(0)$ has a finite value). Since the  lattice 
regularization has no impact on the continuous long distance properties 
of the considered system this term will be just omitted in our further study.

Introducing the $N$-component scalar field replica Hamiltonian
\begin{equation}
\label{2.4}
   H_{N}[{\boldsymbol \phi}] =  
   \frac{1}{2} \int_{0}^{L} d\tau \Biggl(
   \sum_{a=1}^{N} \bigl[\partial_\tau\phi_{a}(\tau)\bigr]^2 
   - \beta u \sum_{a\not= b}^{N} 
   \delta\bigl[\phi_{a}(\tau)-\phi_{b}(\tau)\bigr] \Biggr)
\end{equation}
for the replica partition function, Eq.(\ref{2.3}), 
we obtain the standard expression
\begin{equation}
   \label{2.5}
   Z(N,L) = \prod_{a=1}^{N} \int_{\phi_{a}(0)=0}^{\phi_{a}(L)=0} 
   {\cal D} \phi_{a}(\tau) \;
   \mbox{\Large e}^{-\beta H_{N}[{\boldsymbol \phi}] }
\end{equation}
where ${\boldsymbol \phi} \equiv \{\phi_{1},\dots, \phi_{N}\}$.
According to the above definition this partition function describes the statistics
of $N$ $\delta$-interacting (attracting) trajectories $\phi_{a}(\tau)$ all starting 
(at $\tau=0$) and ending (at $\tau=L$) at zero: $\phi_{a}(0) = \phi_{a}(L) = 0$

In order to map the problem to one-dimensional quantum bosons, instead of the 
above replica partition function, eq.(\ref{2.5}),
 let us introduce more general object
\begin{equation}
   \label{2.6}
   \Psi({\bf x}; t) = 
\prod_{a=1}^{N} \int_{\phi_a(0)=0}^{\phi_a(t)=x_a} {\cal D} \phi_a(\tau)
  \;  \mbox{\Large e}^{-\beta H_{N} [{\boldsymbol \phi}]}
\end{equation}
which describes $N$ trajectories $\phi_{a}(\tau)$ all starting at zero ($\phi_{a}(0) = 0$),
but ending at $\tau = t$ in arbitrary given points $\{x_{1}, ..., x_{N}\}$.
One can easily show that instead of using the path integral, $\Psi({\bf x}; t)$
may be obtained as the solution of the  linear differential equation
\begin{equation}
   \label{2.7}
\partial_t \Psi({\bf x}; t) \; = \;
\frac{1}{2\beta}\sum_{a=1}^{N}\partial_{x_a}^2 \Psi({\bf x}; t)
  \; + \; \frac{1}{2}\beta^2 u \sum_{a\not=b}^{N} \delta(x_a-x_b) \Psi({\bf x}; t)
\end{equation}
with the initial condition 
\begin{equation}
   \label{2.8}
\Psi({\bf x}; 0) = \Pi_{a=1}^{N} \delta(x_a)
\end{equation}
One can easily see that Eq.(\ref{2.7}) is the imaginary-time
Schr\"odinger equation
\begin{equation}
   \label{2.9}
-\partial_t \Psi({\bf x}; t) = \hat{H} \Psi({\bf x}; t)
\end{equation}
with the Hamiltonian
\begin{equation}
   \label{2.10}
   \hat{H} = 
    -\frac{1}{2\beta}\sum_{a=1}^{N}\partial_{x_a}^2 
   -\frac{1}{2}\beta^2 u \sum_{a\not=b}^{N} \delta(x_a-x_b) 
\end{equation}
which describes  $N$ bose-particles of mass $\beta $ interacting via
the {\it attractive} two-body potential $-\beta^2 u \delta(x)$. 
The original replica partition function, Eq.(\ref{2.5}), then is obtained via a particular
choice of the final-point coordinates,
\begin{equation}
   \label{2.11}
   Z(N,L) = \Psi({\bf 0};L).
\end{equation}

%%%%%%%%%%%%%%%%%%%%%%%%%%%%%%%%%%%%%%%%%%%%%%%%%%%%%%%%%%%%%%%%%

Historically the main interest in the studies of such type of systems
was devoted to the quantum bosons with {\it repulsion}. It is for the
case of repulsive interactions the free energy of the $N$ particle system
reveals "correct" extensive behavior, $F \propto N$.
 The eigenfunction of $N$-particle Hamiltonian 
 Eq.(\ref{2.10}), with repulsive interactions, $u < 0$, have been derived 
by Lieb and Liniger in 1963 \cite{Lieb-Liniger}
(for details see Appendix A, as well as Refs. \cite{bogolubov, gaudin}). 
The system of {\it attractive} bosons remained much less studied. 
The free energy of such system reveals "bad" thermodynamic limit 
behavior, namely $F \propto -N^{3}$. Besides, as will be shown later,
the structure of the eigenstates of such system is much more complicated
compared to the case of repulsion.
The spectrum and some properties of the eigenfunctions
for  attractive ($u > 0$) one-dimensional quantum boson system have been derived by McGuire
\cite{McGuire} and by Yang \cite{Yang} (see also Ref.\ \cite{Takahashi,Calabrese}).
Detailed structure and the properties of these wave functions are described in 
Appendix B. 

A generic eigenstate of such system consists of $M$  ($1 \leq M \leq N$)  
"clusters" $\{\Omega_{\alpha}\}$ $(\alpha = 1,...,M)$ of bound particles. 
Each cluster is characterized by the momentum $q_{\alpha}$
of its center of mass motion, and by the number $n_{\alpha}$ of particles contained in it
(such that $\sum_{\alpha=1}^{M} n_{\alpha} = N$).  
Correspondingly, the eigenfunction $\Psi_{\bf q, n}^{(M)}(x_1,...,x_N)$ of such state
is characterized by $M$ continuous parameters ${\bf q} = (q_{1}, ..., q_{M})$  and $M$ 
integer parameters ${\bf n} = (n_{1}, ..., n_{M})$ (see Appendix B2, Eq.(\ref{B27})). 
The energy spectrum of this state is
\begin{equation}
\label{2.12}
E_{M}({\bf q,n})  = 
 \frac{1}{2\beta} \sum_{\alpha=1}^{M} \; n_{\alpha} q_{\alpha}^{2} 
- \frac{\kappa^{2}}{24\beta}\sum_{\alpha=1}^{M} (n_{\alpha}^{3}-n_{\alpha})
\end{equation}
where 
\begin{equation}
\label{2.13}
\kappa \; = \; \beta^{3} u
\end{equation}
A general time dependent solution  $\Psi({\bf x},t)$
of the Schr\"odinger equation (\ref{2.7}) with the initial conditions, Eq.(\ref{2.8}),
can be represented in the form of the linear combination of the eigenfunctions
$\Psi_{\bf q, n}^{(M)}({\bf x})$ (Appendix B4, Eq.(\ref{B50})):
\begin{equation}
\label{2.14}
\Psi({\bf x},t) \; = \; \sum_{M=1}^{N} \;
{\sum_{{\bf n}}}' \; 
\int ' {\cal D} {\bf q} \; \; 
 \Psi_{\bf q, n}^{(M)}({\bf x}) \Psi_{\bf q, n}^{(M)^{*}}({\bf 0}) \; 
\exp\bigl[-E_{M}({\bf q,n}) \; t \bigr]
\end{equation}
where the summations over the integer parameters $n_{\alpha}$ and the integrations over the 
momenta $q_{\alpha}$ are performed in a restricted subspace, Eqs.(\ref{B42})-(\ref{B45}) and (\ref{B51}),  
which reflects the specific symmetry properties of the eigenfunctions 
$\Psi_{\bf q, n}^{(M)}({\bf x})$.
Correspondingly, according to Eq.(\ref{2.11}), for the replica partition function 
of the original directed polymer problem one gets (Appendix B4, Eq.(\ref{B59}))
\begin{equation}
\label{2.15}
Z(N,L) \; = \;  
\sum_{M=1}^{N} \; 
\frac{1}{M!}
\biggl[
\prod_{\alpha=1}^{M} 
\int_{-\infty}^{+\infty} \; \frac{dq_{\alpha}}{2\pi} 
\sum_{n_{\alpha}=1}^{\infty}
\biggr] \;
{\boldsymbol \delta}\biggl(\sum_{\alpha=1}^{M} n_{\alpha}, \; N \biggr) \;
\big| \Psi_{\bf q, n}^{(M)}({\bf 0}) \big|^2 \; 
\mbox{\LARGE e}^{-E_{M}({\bf q,n}) L}
\end{equation}
where due to the symmetry of the function 
$f\bigl({\bf q}, {\bf n}\bigr) = \bigl| \Psi_{\bf q, n}^{(M)}({\bf 0}) \bigr|^2 \; 
\exp\bigl[-E_{M}({\bf q,n}) \; L \bigr]$
with respect to permutations of all its $M$ pairs of arguments $(q_{\alpha}, n_{\alpha})$
the integrations over $M$ momenta $q_{\alpha}$ can be extended to the whole space $R_{M}$
while the summations over  $n_{\alpha}$'s are bounded by the only constrain
$\sum_{\alpha=1}^{M} n_{\alpha} = N$ (for simplicity, due to the presence
of the Kronecker symbol ${\boldsymbol \delta}\bigl(\sum_{\alpha} n_{\alpha}, \; N \bigr)$,
the summations over $n_{\alpha}$'s are extended to infinity).

\section{Solution of the one-dimensional directed polymers problem}

\newcounter{3}
\setcounter{equation}{0}
\renewcommand{\theequation}{3.\arabic{equation}}

Using the explicit form of the wave functions $\Psi_{\bf q, n}^{(M)}({\bf x})$, Eq.(\ref{B27}),
the expression in Eq.(\ref{2.15}) for the replica partition function
can be reduced to (Appendix B4, Eq.(\ref{B61})-(\ref{B62}))
\begin{equation}
   \label{3.1}
 Z(N,L) \; =  \mbox{\LARGE e}^{-\beta N L f_{0}} \;\;
\tilde{Z}(N,\lambda) 
\end{equation}
where $f_{0} = \frac{1}{24}\beta^4 u^2 - \frac{1}{\beta L} \ln(\beta^{3} u)$ 
is the linear (selfaveraging) free energy density (cf. Eq.(\ref{1.19})), and
\begin{eqnarray}
\nonumber
\tilde{Z}(N.L) &=& 
N! \; 
 \int_{-\infty}^{+\infty} \frac{dq}{2\pi\kappa N} \;
\exp\Bigl[
-\frac{L}{2\beta} N q^{2} + \frac{\kappa^{2}L}{24\beta} N^{3} 
\Bigr]
\; +
\\
\nonumber
\\
\nonumber
&+& N! \; 
\sum_{M=2}^{N} \frac{1}{M!} 
\biggl[
\prod_{\alpha=1}^{M}
\sum_{n_{\alpha}=1}^{\infty}
\int_{-\infty}^{+\infty} \frac{d q_{\alpha}}{2\pi\kappa n_{\alpha}} 
\biggr]
\;{\boldsymbol \delta}\biggl(\sum_{\alpha=1}^{M} n_{\alpha}, \; N \biggr) 
\prod_{\alpha<\beta}^{M} 
\frac{\big|q_{\alpha}-q_{\beta} -\frac{i\kappa}{2}(n_{\alpha}-n_{\beta})\big|^{2}}{
      \big|q_{\alpha}-q_{\beta} -\frac{i\kappa}{2}(n_{\alpha}+n_{\beta})\big|^{2}} 
\times
\\
&\times&
\exp\Bigl[
-\frac{L}{2\beta}\sum_{\alpha=1}^{M} n_{\alpha} q_{\alpha}^{2} + 
\frac{\kappa^{2}L}{24\beta} \sum_{\alpha=1}^{M} n_{\alpha}^{3}
\Bigr] 
\label{3.2}
\end{eqnarray}
The first term in the above expression is the contribution of the ground state $(M=1)$,
while  the next terms $(M \geq 2)$ are the contributions of the rest of the energy
spectrum.

The terms cubic in $n_{\alpha}$ in the exponential of Eq.\ (\ref{3.2})
can be linearised with the help of Airy function, using the standard relation 
(see Appendix C)
\begin{equation}
   \label{3.3}
\exp\Bigl( \frac{1}{3} \lambda^{3} n^{3}\Bigr) \; = \; 
\int_{-\infty}^{+\infty} dy \; \Ai(y) \; \exp(\lambda n)
\end{equation}
Redefining the momenta, $q_{\alpha} = \bigl(\beta\kappa/L\bigr)^{1/3} p_{\alpha}$ 
and introducing a new parameter
\begin{equation}
   \label{3.4}
\lambda (L) \; = \; \frac{1}{2} \biggl(\frac{L}{\beta} \kappa^{2} \biggr)^{1/3} \; = \; 
\frac{1}{2} \Bigl(\beta^{5} u^{2} L \Bigr)^{1/3}
\end{equation}
we get
\begin{eqnarray}
\label{3.5a}
\tilde{Z}(N,\lambda) &=& N! 
\int\int_{-\infty}^{+\infty} \frac{dy dp}{4\pi\lambda N} 
\; \Ai(y) \; \mbox{\LARGE e}^{\lambda N (y-p^{2})}  + 
\\
\nonumber
 &+& N! \sum_{M=2}^{N} \frac{1}{M!} 
\biggl[\prod_{\alpha=1}^{M} \sum_{n_{\alpha}=1}^{\infty} \int\int_{-\infty}^{+\infty}
 \frac{dy_{\alpha} dp_{\alpha}}{4\pi\lambda n_{\alpha}} 
\Ai(y_{\alpha}) \; \mbox{\LARGE e}^{\lambda n_{\alpha} (y_{\alpha}-p^{2}_{\alpha})}\biggr] 
\prod_{\alpha<\beta}^{M}
\frac{\big|\lambda(n_{\alpha}-n_{\beta}) -i(p_{\alpha}-p_{\beta})\big|^{2}}{
      \big|\lambda(n_{\alpha}+n_{\beta}) -i(p_{\alpha}-p_{\beta}) \big|^{2}} \;
{\boldsymbol \delta}\biggl(\sum_{\alpha=1}^{M}n_{\alpha}, \; N\biggr)
\end{eqnarray}
After shifting the Airy function parameters of integration
$y_{\alpha} \to y_{\alpha}  + p_{\alpha}^{2}$ the expression 
for $\tilde{Z}(N,\lambda)$ becomes sufficiently compact:
\begin{eqnarray}
\label{3.5}
\tilde{Z}(N,\lambda) &=& N! 
\int\int_{-\infty}^{+\infty} \frac{dy dp}{4\pi\lambda N} 
\; \Ai(y+p^{2}) \; \mbox{\LARGE e}^{\lambda N y}  + 
\\
\nonumber
 &+& N! \sum_{M=2}^{N} \frac{1}{M!} 
\biggl[\prod_{\alpha=1}^{M} \sum_{n_{\alpha}=1}^{\infty} \int\int_{-\infty}^{+\infty}
 \frac{dy_{\alpha} dp_{\alpha}}{4\pi\lambda n_{\alpha}} 
\Ai(y_{\alpha}+p^{2}_{\alpha}) \; \mbox{\LARGE e}^{\lambda n_{\alpha} y_{\alpha}}\biggr] 
\prod_{\alpha<\beta}^{M}
\frac{\big|\lambda(n_{\alpha}-n_{\beta}) -i(p_{\alpha}-p_{\beta})\big|^{2}}{
      \big|\lambda(n_{\alpha}+n_{\beta}) -i(p_{\alpha}-p_{\beta}) \big|^{2}} \;
{\boldsymbol \delta}\biggl(\sum_{\alpha=1}^{M}n_{\alpha}, \; N\biggr)
\end{eqnarray}
Now, using the {\it Cauchy double alternant identity}
\begin{equation}
 \label{3.6}
\frac{\prod_{\alpha<\beta}^{M} (a_{\alpha} - a_{\beta})(b_{\alpha} - b_{\beta})}{
     \prod_{\alpha,\beta=1}^{M} (a_{\alpha} - b_{\beta})} \; = \; 
(-1)^{M(M-1)/2} \det\Bigl[\frac{1}{a_{\alpha}-b_{\beta}}\Bigr]_{\alpha,\beta=1,...M}
\end{equation}
and introducing 
$a_{\alpha} = p_{\alpha} - i \lambda n_{\alpha}$ and 
$b_{\alpha} = p_{\alpha} + i \lambda n_{\alpha}$, 
the product term in eq.(\ref{3.5}) can be represented in the determinant form:
\begin{equation}
\prod_{\alpha<\beta}^{M}
\frac{\big|\lambda(n_{\alpha}-n_{\beta}) -i(p_{\alpha}-p_{\beta})\big|^{2}}{
      \big|\lambda(n_{\alpha}+n_{\beta}) -i(p_{\alpha}-p_{\beta}) \big|^{2}}
 =  
\biggl[\prod_{\alpha=1}^{M} (2\lambda n_{\alpha}) \biggr]\; 
\det\Bigl[\frac{1}{\lambda n_{\alpha} -ip_{\alpha} 
                 + \lambda n_{\beta} + ip_{\beta}}\Bigr]_{\alpha,\beta=1,...M}
\label{3.7}
\end{equation}
Substituting now the expression for the replica partition function $\tilde{Z}(N,\lambda)$
into the definition of the probability function, eq.(\ref{1.24}), we can perform summation over $N$
(which would lift the constraint $\sum_{\alpha=1}^{M}n_{\alpha} =  N$) and obtain:
\begin{equation}
 \label{3.8}
W(x)  = 
\lim_{\lambda\to\infty} \Biggl\{
1  +  \sum_{M=1}^{\infty} \frac{(-1)^{M}}{M!}
\Biggl[\prod_{\alpha=1}^{M}
\int\int_{-\infty}^{+\infty}
\frac{dy_{\alpha} dp_{\alpha}}{2\pi}  
\Ai(y_{\alpha}+p^{2}_{\alpha})
\sum_{n_{\alpha}=1}^{\infty} (-1)^{n_{\alpha}-1} \mbox{\LARGE e}^{\lambda n_{\alpha} (y_{\alpha}+x)}
\Biggr] 
\det\Bigl[\frac{1}{\lambda n_{\alpha} -ip_{\alpha} + \lambda n_{\beta} + ip_{\beta}}\Bigr] \Biggr\}
\end{equation}
The above expression in nothing else but the expansion of the Fredholm determinant 
$\det(1 - \hat{K})$
(see e.g. \cite{Mehta}, Appendix D)
with the kernel
\begin{equation}
\label{3.9}
\hat{K} \equiv
K\bigl[(n,p); (n',p')\bigr] = 
\Biggl[\int_{-\infty}^{+\infty} dy \Ai(y+p^{2}) (-1)^{n-1} \mbox{\LARGE e}^{\lambda n (y+x)}\Biggr]
\frac{1}{\lambda n -ip + \lambda n' + ip'} 
\end{equation}
Using the exponential representation of this determinant we get
\begin{equation}
 \label{3.10}
W(x)  = 
\lim_{\lambda\to\infty}
\exp\Bigl[-\sum_{M=1}^{\infty} \frac{1}{M} \; \mbox{Tr} \hat{K}^{M} \Bigr]
\end{equation}
where 
\begin{eqnarray}
\nonumber
\mbox{Tr} \hat{K}^{M} &=& 
\Biggl[\prod_{\alpha=1}^{M}
\int\int_{-\infty}^{+\infty}
\frac{dy_{\alpha} dp_{\alpha}}{2\pi}  
\Ai(y_{\alpha}+p^{2}_{\alpha})
\sum_{n_{\alpha}=1}^{\infty} (-1)^{n_{\alpha}-1} \mbox{\LARGE e}^{\lambda n_{\alpha} (y_{\alpha}+x)}
\Biggr] \times
\\
\nonumber
\\
&\times&
\frac{1}{(\lambda n_{1} -ip_{1} + \lambda n_{2} + ip_{2})
         (\lambda n_{2} -ip_{2} + \lambda n_{3} + ip_{3})...
(\lambda n_{M} -ip_{M} + \lambda n_{1} + ip_{1})}
 \label{3.11}
\end{eqnarray}
Substituting here
\begin{equation}
 \label{3.12}
\frac{1}{\lambda n_{\alpha} -ip_{\alpha} + \lambda n_{\alpha+1} + ip_{\alpha+1}}  =  
\int_{0}^{\infty} d\omega_{\alpha} 
\exp\bigl[-( \lambda n_{\alpha} - ip_{\alpha} + \lambda n_{\alpha+1} + ip_{\alpha+1}) \omega_{\alpha} \bigr]
\end{equation}
one can easily perform the summation over $n_{\alpha}$'s. Taking into account that
\begin{equation}
 \label{3.13}
\lim_{\lambda\to\infty} 
\sum_{n=1}^{\infty} (-1)^{n-1} \mbox{\LARGE e}^{\lambda n z} \; = \; 
\lim_{\lambda\to\infty} \frac{\mbox{\LARGE e}^{\lambda z}}{1 + \mbox{\LARGE e}^{\lambda z}}  =  
\theta(z)
\end{equation}
we get
\begin{equation}
 \label{3.14a} 
\lim_{\lambda\to\infty} Tr \; \hat{K}^{M} = 
\prod_{\alpha=1}^{M}
\int\int_{-\infty}^{+\infty} \frac{dy_{\alpha} dp_{\alpha}}{2\pi} 
\int_{0}^{\infty} d\omega_{\alpha} 
\Ai(y_{\alpha}+p^{2}_{\alpha})
\theta(y_{\alpha} + x -\omega_{\alpha} - \omega_{\alpha-1})
\mbox{\LARGE e}^{ip_{\alpha} (\omega_{\alpha}-\omega_{\alpha-1})}
\end{equation}
where by definition  $\omega_{0} \equiv \omega_{M}$. 
Shifting the integration parameters, $y_{\alpha} \to y_{\alpha} - x + \omega_{\alpha} + \omega_{\alpha-1}$
and $\omega_{\alpha} \to  \omega_{\alpha} + x/2$, 
we obtain
\begin{equation}
\lim_{\lambda\to\infty} Tr \; \hat{K}^{M} = 
\prod_{\alpha=1}^{M}
\int_{0}^{\infty} dy_{\alpha} 
\int_{-\infty}^{+\infty} \frac{dp_{\alpha}}{2\pi}  
\int_{-x/2}^{\infty} d\omega_{\alpha} 
\Ai(y_{\alpha}+p^{2}_{\alpha}+\omega_{\alpha}+\omega_{\alpha-1}) \; 
\mbox{\LARGE e}^{ip_{\alpha} (\omega_{\alpha}-\omega_{\alpha-1})}
 \label{3.14}
\end{equation}
Using the Airy function integral representation,
and taking into account that it satisfies the differential equation, $\Ai''(t) = t \Ai(t)$, 
one can easily perform the following integrations (see Appendix C):
\begin{eqnarray}
\int_{0}^{\infty} dy 
\int_{-\infty}^{+\infty} \frac{dp}{2\pi}  
\Ai(y + p^{2} + \omega + \omega') 
\mbox{\LARGE e}^{ip (\omega-\omega')} &=&
 2^{-1/3} \int_{0}^{\infty} dy 
\Ai\bigl(2^{1/3} \omega + y\bigr) 
\Ai\bigl(2^{1/3} \omega' + y\bigr) 
\\
\nonumber
&=&
\frac{\Ai\bigl(2^{1/3} \omega\bigr) \Ai'\bigl(2^{1/3} \omega'\bigr) - 
      \Ai'\bigl(2^{1/3} \omega\bigr) \Ai\bigl(2^{1/3} \omega'\bigr)}{
\omega - \omega'}
\label{3.15}
\end{eqnarray}
Redefining $\omega_{\alpha} \to \omega_{\alpha} 2^{-1/3}$ we find
\begin{equation}
\lim_{\lambda\to\infty} Tr \hat{K}^{M} = 
\int\int ...\int_{-x/2^{2/3}}^{\infty} d\omega_{1} d\omega_{2} ... d\omega_{M}
K_{A}(\omega_{1},\omega_{2}) K_{A}(\omega_{2},\omega_{3}) ... K_{A}(\omega_{M},\omega_{1})
\label{3.16}
\end{equation}
where
\begin{equation}
 \label{3.17}
K_{A}(\omega,\omega') \; = \; 
\frac{\Ai(\omega) \Ai'(\omega') - \Ai'(\omega) \Ai(\omega')}{
\omega - \omega'}
\end{equation}
is the so called Airy kernel. This proves that in the thermodynamic limit, $L \to \infty$,
the probability function $W(x)$, eq.(\ref{1.22}), is defined by the Fredholm determinant,
\begin{equation}
 \label{3.18}
W(x)  \; = \; \det[1 - \hat{K}_{A}] \equiv F_{2}(-x/2^{2/3})
\end{equation}
where $\hat{K}_{A}$ is the integral operator on $[-x/2^{2/3}, \infty)$ with the 
Airy kernel, eq.(\ref{3.17}). This is  the Tracy-Widom distribution function 
which has the following explicit form (see Appendix D):
\begin{equation}
 \label{3.19}
F_{2}(s) \; = \; \exp\Bigl(-\int_{s}^{\infty} dt \; (t-s) \; q^{2}(t)\Bigr)
\end{equation}
where the function $q(t)$ is the solution of the Panlev\'e II equation, 
$q'' = t q + 2 q^{3}$
with the boundary condition, $q(t\to +\infty) \sim Ai(t)$. 
Note that according to Eqs.(\ref{1.22}), (\ref{1.32}) and (\ref{3.18}),
$P_{*}(x) = 2^{-2/3} P_{TW}(-2^{-2/3} x)$

\section{Conclusions}

The first breakthrough in the studies of one-dimensional directed polymers
in random potential was due to the work of Kardar \cite{kardar_87},
in which the problem has been reduced to $N$-particle system of quantum bosons
with attractive interactions. By that time the very first idea was that 
in the thermodynamic limit it would be sufficient to take into account 
only the contribution of the ground state whose energy was well known.
Indeed, for any integer $N > 1$ the contribution of the excited states
in the limit $L \to \infty$ are exponentially small compared to that 
of the ground state. In the framework of this approach it has been 
demonstrated that the free energy fluctuations grow as $L^{1/3}$
while the typical value of the polymers deviations scale as $L^{2/3}$
which, in particular, was in perfect agreement with numerical studies.

However, more detailed investigations demonstrated that the above approach
reveals serious pathologies. In particular it turned out that the
second cummulant of the free energy $\bigl( \overline{F^{2}} - \overline{F}^{2}\bigr)$
appears to be identically equal to zero! This is possible only in two
cases: either the quantity $F$ is not random (which contradicts to the
fact that its fluctuations scale as $L^{1/3}$), or the distribution function of the
this quantity is not positively defined (which, of course, makes no physical
sense). Simple mathematical analysis demonstrated that the origin of this
pathology is hidden in the replicas "magic operations": on one hand, all the 
calculations are performed assuming that the replica parameter $N$ (number
of particles) is an integer $N > 1$, while on the other hand, in the thermodynamic limit
$L \to \infty$ the relevant values of the parameter $N$ which defines the 
physical properties of the original random system appears to be in the region
$N \to 0$. In other words, the replica method assumes {\it analytic continuation}
of the result obtained for arbitrary integers $N$ to the region $N \to 0$.
The problem is that, first, such analytic continuation is not always unambiguous
(see e.g.  \cite{REM,replicas}), and second, any approximations in the 
calculations of the integer-valued replica partition function are quite risky
for the validity of the further analytic continuation to non-integer values
of $N$. 

For the problem under consideration the point is that, in fact, neglected
 exponentially small contributions at integers $N > 1$ appear to be quite
essential in the region $N\to 0$, which defines the properties of the
free energy distribution function $P_{*}(f) = \lim_{L\to \infty} P_{L}(f) $.
In other words, the problem is that the two limits, $L\to\infty$ and  $N\to 0$,
do not commute \cite{Medina_93,dirpoly}.

Nevertheless,  in terms of this approximation (assuming the universal scaling $L^{1/3}$ 
of the free energy fluctuations)
 one can derive the  left tail of the distribution function
$P_{*}(f)$ which is given by the asymptotics of the Airy function,
$P_{*}(f\to -\infty) \sim \exp\bigl(-\frac{2}{3}\big|f\big|^{3/2}\bigr)$ \cite{Zhang}.
For the first time the form of the {\it right tail} of this distribution function,
$P_{*}(f\to +\infty) \sim \exp\bigl[-(const) f^{3}\bigr]$,
has been derived in terms of the optimal fluctuation approach 
\cite{KK1,KK2,KK3}, where it has been 
demonstrated that both asymptotics (left and right) of the 
function $P_{*}(f)$ are consistent with the Tracy-Widom 
distribution \cite{Tracy-Widom} which was known to describe the
statistical properties of many other systems 
\cite{PNG_Spohn,LIS,LCS,oriented_boiling,ballistic_decomposition,DP_johansson}.

For the first time TW distribution for the directed polymers
with $\delta$-correlated random potentials was derived in terms of the
distribution of the solutions of the KPZ equation \cite{KPZ-TW1,KPZ-TW2},
which, in particular, describes the domain walls growth, and which is equivalent
to the present system. Almost simultaneously the exact solution 
of the one-dimensional directed polymer problem has been found in terms of 
of the replica method, which involved the summation over the whole
spectrum of excited states in the corresponding $N$-particle quantum
boson system \cite{Dotsenko1,Dotsenko2,Dotsenko3,LeDoussal}.
These calculations resulted in the derivation of the 
entire free energy distribution, which was proved to coincide with
TW distribution function.

The Tracy-Widom function, Eq.(\ref{3.19}), was originally derived 
for rather specific mathematical problem, namely for the probability 
distribution of the
largest eigenvalue of a $N\times N$ random hermitian matrix in the 
limit $N \to \infty$ \cite{Tracy-Widom}.
It is amazing but now there are exists a long  list of statistical systems 
(which at first sight have just nothing to do with the original 
random matrix problem) 
whose macroscopic properties are described 
by {\it the same} universal TW distribution function. 
In other words, all the above observations indicate there exist
a kind of "superuniversality" for the entire class of various random 
systems.

In this review we have described the exact solution of the problem
which remained unsolved during last almost thirty years. 
It should be stressed that this solution has been obtained in terms of the
replica method. This is very rare case when the solution of a non-trivial
problem has been found without using heuristic "replica magic" operations,
quite typical for this method, which usually forced to think that the 
"replica method" and the "exact solution" are two things absolutely incompatible.
Hopefully finding exact solution is not always means the end of the
story: methodology and created mathematical technique could be used
for solving numerous other problems still waiting for their solutions...

%%%%%%%%%%%%%%%%%%%%%%%%%%%%%%%%%%%%%%

%\newpage

\vspace{5mm}

\begin{center}

\appendix{\Large Appendix A}

\vspace{5mm}

 {\bf \large Quantum bosons with repulsive interactions}

\end{center}

\newcounter{A}
\setcounter{equation}{0}
\renewcommand{\theequation}{A.\arabic{equation}}

\vspace{5mm}

{\center{\bf 1. Eigenfunctions}}

\vspace{3mm}

The eigenstates equation 
for  $N$-particle system of one-dimensional quantum bosons with $\delta$-interactions is
\begin{equation}
 \label{A1}
\frac{1}{2}\sum_{a=1}^{N}\partial_{x_a}^2 \Psi({\bf x}) \; + \; 
\frac{1}{2}\kappa \sum_{a\not=b}^{N} \delta(x_a-x_b) \Psi({\bf x}) 
\; = \; - \beta E \Psi({\bf x})
\end{equation}
(where $\kappa = \beta^{3} u$). Due to the symmetry of the wave function 
with respect to permutations of its arguments
it is sufficient to consider it in the sector 
\begin{equation}
 \label{A2}
x_{1} \; < \; x_{2} \; < \; ... \; < \; x_{N}
\end{equation}
as well as at its boundary. Inside this sector the wave function $\Psi({\bf x})$ 
satisfy the equation
\begin{equation}
 \label{A3}
\frac{1}{2}\sum_{a=1}^{N}\partial_{x_a}^2 \Psi({\bf x}) 
\; = \; - \beta E \Psi({\bf x})
\end{equation}
which describes $N$ free particles, and its generic solution is the linear combination
of $N$ plane waves characterized by $N$ momenta $\{q_{1}, q_{2}, ..., q_{N}\} \equiv {\bf q}$.
Integrating Eq.(\ref{A1}) over the variable $(x_{i+1}-x_{i})$ in a small interval
around zero, $|x_{i+1}-x_{i}| < \epsilon \to  0$, and assuming that the other
variables $\{ x_{j}\}$ (with $j \not= i, i+1$) belong to the sector, Eq.(\ref{A2}),  
one easily finds that the wave function  $\Psi({\bf x})$ must satisfy the following
boundary conditions:
\begin{equation}
\label{A4}
\bigl(\partial_{x_{i+1}} - \partial_{x_i} + \kappa \bigr) 
\Psi({\bf x})\bigg|_{x_{i+1} = x_{i} + 0} \; = \; 0
\end{equation}
Functions satisfying both Eq.\ (\ref{A3}) and
the boundary conditions Eq.\ (\ref{A4}) can be written in the form
\begin{equation}
\label{A5}
\Psi_{q_{1}...q_{N}}(x_{1}, ..., x_{N}) \equiv
\Psi^{(N)}_{\bf q}({\bf x}) \; = \; C 
\biggl(\prod_{a<b}^{N}\bigl[ \partial_{x_a} - \partial_{x_b} + \kappa \bigr]\biggr) \;
    \det\Bigl[\exp( i q_c \, x_d) \Bigr]_{(c,d)=1,...,N}
\end{equation}
where $C$ is the normalization constant to be defined later.
First of all, it is evident that being the linear combination of the 
plane waves, the above wave function satisfy  Eq.(\ref{A3}). To demonstrate
which way this function satisfy the boundary conditions, Eq.(\ref{A4}),
let us check it for the case $i=1$.
According to Eq.(\ref{A5}), the wave function $\Psi^{(N)}_{\bf q}({\bf x})$
can be represented in the form
\begin{equation}
\label{A6}
\Psi^{(N)}_{\bf q}({\bf x}) \; = \; 
- \bigl(\partial_{x_2} - \partial_{x_1} - \kappa \bigr) 
\tilde{\Psi}^{(N)}_{\bf q}({\bf x})
\end{equation}
where
\begin{equation}
\label{A7}
\tilde{\Psi}^{(N)}_{\bf q}({\bf x}) \; = \; C 
\biggl(\prod_{a=3}^{N}\bigl[ \partial_{x_1} - \partial_{x_a} + \kappa \bigr]
                      \bigl[ \partial_{x_2} - \partial_{x_a} + \kappa \bigr] \biggr)
\biggl(\prod_{3\leq a <b}^{N}\bigl[ \partial_{x_a} - \partial_{x_b} + \kappa \bigr]\biggr)
    \det\Bigl[\exp( i q_c \, x_d) \Bigr]_{(c,d)=1,...,N}
\end{equation}
One can easily see that this function is {\it antisymmetric} with respect to
the permutation of $x_{1}$ and $x_{2}$. Substituting Eq.(\ref{A6}) 
into Eq.(\ref{A4}) (with $i=1$)
we get
\begin{equation}
\label{A8}
-\biggl[\bigl(\partial_{x_2} - \partial_{x_1}\bigr)^{2} - \kappa^{2} \biggr] 
\tilde{\Psi}^{(N)}_{\bf q}({\bf x})\bigg|_{x_{2} = x_{1}} \; = \; 0
\end{equation}
Given the antisymmetry of the l.h.s expression with respect to
the permutation of $x_{1}$ and $x_{2}$ the above condition is indeed satisfied
at boundary $x_{1}=x_{2}$.

Since the eigenfunction $\Psi^{(N)}_{\bf q}({\bf x})$ satisfying 
Eq.(\ref{A1}) must be {\it symmetric} with respect to permutations of
its arguments, the function, Eq.(\ref{A5}), can be easily continued 
beyond the sector, Eq.(\ref{A2}), to the entire space of variables
$\{x_{1}, x_{2}, ..., x_{N}\} \in R_{N}$,
\begin{equation}
\label{A9}
\Psi^{(N)}_{\bf q}({\bf x}) \; = \; C 
\biggl(\prod_{a<b}^{N}\Bigl[ -i\bigl(\partial_{x_a} - \partial_{x_b}\bigr) +i \kappa \sgn(x_{a}-x_{b})\Bigr]\biggr) \;
    \det\Bigl[\exp( i  q_c \, x_d) \Bigr]_{(c,d)=1,...,N}
\end{equation}
where, by definition, the differential operators $\partial_{x_a}$ act only on 
the exponential terms and not on the $\sgn(x)$ functions, and 
for further convenience we have redefined $i^{N(N-1)/2} C \to C$. 
Explicitly the determinant in the above equation is
\begin{equation}
\label{A10}
     \det\Bigl[\exp( i  q_c \, x_d) \Bigr]_{(c,d)=1,...,N} \; = \; 
\sum_{P} (-1)^{[P]} \; \exp\Bigl[i \sum_{a=1}^{N} q_{p_{a}} x_{a} \Bigr]
\end{equation}
where the summation goes over the permutations $P$ of $N$ momenta $\{ q_{1}, q_{2}, ..., q_{N}\}$
over $N$ particles $\{ x_{1}, x_{2}, ..., x_{N}\}$, and $[P]$ denotes the parity of the permutation.
In this way the eigenfunction, Eq.(\ref{A9}), can be represented as follows
\begin{equation}
\label{A11}
\Psi^{(N)}_{\bf q}({\bf x}) \; = \; C 
\sum_{P} (-1)^{[P]} \; 
\biggl(\prod_{a<b}^{N}\Bigl[ -i\bigl(\partial_{x_a} - \partial_{x_b}\bigr) +i \kappa \sgn(x_{a}-x_{b})\Bigr]\biggr) \;
\exp\Bigl[i \sum_{a=1}^{N} q_{p_{a}} x_{a} \Bigr]
\end{equation}
 Taking the derivatives, we obtain
\begin{equation}
\label{A12}
\Psi^{(N)}_{\bf q}({\bf x}) \; = \; C 
\sum_{P} (-1)^{[P]} \; 
\biggl(\prod_{a<b}^{N}\Bigl[ q_{p_a} - q_{p_b} +i \kappa \sgn(x_{a}-x_{b})\Bigr]\biggr) \;
\exp\Bigl[i \sum_{a=1}^{N} q_{p_{a}} x_{a} \Bigr]
\end{equation}
It is evident from these representations that the eigenfunctions 
$\Psi^{(N)}_{\bf q}({\bf x})$ are
{\it antisymmetric} with respect to permutations of the momenta $q_{1}, ..., q_{N}$. 

Finally, substituting the expression for the eigenfunctions, Eq.(\ref{A5}) (which is valid in
the sector, Eq.(\ref{A2})), into Eq.(\ref{A3}) for the energy spectrum we find
\begin{equation}
   \label{A13}
E \;  = \; \frac{1}{2\beta} \sum_{a=1}^{N} q_{a}^{2} 
\end{equation}

\vspace{10mm}

{\center{\bf 2. Orthonormality}}

\vspace{3mm}

Now one can easily prove that the above eigenfunctions with different momenta are orthogonal to
each other. Let us consider two wave functions $\Psi^{(N)}_{\bf q}({\bf x})$ and 
$\Psi^{(N)}_{\bf q'}({\bf x})$ where it is assumed that
\begin{eqnarray}
 \label{A14}
q_{1} \; < \; q_{2} \; < \; ... \; < \; q_{N}\\
\nonumber
q'_{1} \; < \; q'_{2} \; < \; ... \; < \; q'_{N}
\end{eqnarray}
Using the representation, Eq.(\ref{A11}), for the overlap of these two function we get
\begin{eqnarray}
 \label{A15}
\nonumber
\overline{\Psi^{(N)^{*}}_{\bf q'}({\bf x}) \Psi^{(N)}_{\bf q}({\bf x})} &\equiv& 
\int_{-\infty}^{+\infty} d^{N}{\bf x} \; \Psi^{(N)^{*}}_{\bf q'}({\bf x}) \Psi^{(N)}_{\bf q}({\bf x}) \\
\nonumber
&=& |C|^{2} \sum_{P,P'} (-1)^{[P]+[P']}  
\int_{-\infty}^{+\infty} d^{N}{\bf x} 
\biggl\{\biggl(\prod_{a<b}^{N}
\bigl[ i\bigl(\partial_{x_a} - \partial_{x_b}\bigr) -i \kappa \sgn(x_{a}-x_{b})\bigr]\biggr) \;
\exp\bigl[-i \sum_{a=1}^{N} q'_{p'_{a}} x_{a} \bigr]\biggr\} \times \\
&\times& 
\biggl\{\biggl(\prod_{a<b}^{N}
\bigl[ -i\bigl(\partial_{x_a} - \partial_{x_b}\bigr) +i \kappa \sgn(x_{a}-x_{b})\bigr]\biggr) \;
\exp\bigl[i \sum_{a=1}^{N} q_{p_{a}} x_{a} \bigr]\biggr\}
\end{eqnarray}
Integrating by parts we obtain
\begin{eqnarray}
 \label{A16}
\overline{\Psi^{(N)^{*}}_{\bf q'}({\bf x}) \Psi^{(N)}_{\bf q}({\bf x})} &=& |C|^{2}
\sum_{P,P'} (-1)^{[P]+[P']}  
\int_{-\infty}^{+\infty} d^{N}{\bf x} \;
\exp\bigl[-i \sum_{a=1}^{N} q'_{p'_{a}} x_{a} \bigr] \times \\
\nonumber
&\times&
\biggl(\prod_{a<b}^{N}\bigl[-i\bigl(\partial_{x_a} - \partial_{x_b}\bigr) -i \kappa \sgn(x_{a}-x_{b})\bigr] \;
                      \bigl[-i\bigl(\partial_{x_a} - \partial_{x_b}\bigr) +i \kappa \sgn(x_{a}-x_{b})\bigr] \biggr)
\exp\bigl[i \sum_{a=1}^{N} q_{p_{a}} x_{a} \bigr]
\end{eqnarray}
or
\begin{equation}
 \label{A17}
\overline{\Psi^{(N)^{*}}_{\bf q'}({\bf x}) \Psi^{(N)}_{\bf q}({\bf x})} = |C|^{2}
\sum_{P,P'} (-1)^{[P]+[P']}  
\int_{-\infty}^{+\infty} d^{N}{\bf x} 
\exp\bigl[-i \sum_{a=1}^{N} q'_{p'_{a}} x_{a} \bigr]
\biggl(\prod_{a<b}^{N}\bigl[-(\partial_{x_a} - \partial_{x_b})^{2} + \kappa^{2} \bigr] \biggr)
\exp\bigl[i \sum_{a=1}^{N} q_{p_{a}} x_{a} \bigr]
\end{equation}
Taking the derivatives and performing the integrations
we find
\begin{eqnarray}
 \label{A18}
\nonumber
\overline{\Psi^{(N)^{*}}_{\bf q'}({\bf x}) \Psi^{(N)}_{\bf q}({\bf x})} &=& 
|C|^{2} \sum_{P,P'} (-1)^{[P]+[P']}  
\biggl(\prod_{a<b}^{N}\bigl[(q_{p_a} - q_{p_b})^{2} + \kappa^{2} \bigr] \biggr)
\int_{-\infty}^{+\infty} d^{N}{\bf x} \; 
\exp\bigl[i \sum_{a=1}^{N} (q_{p_{a}}-q'_{p'_{a}}) x_{a} \bigr] \\ 
&=& 
|C|^{2} \sum_{P,P'} (-1)^{[P]+[P']} 
\biggl(\prod_{a<b}^{N}\bigl[(q_{p_a} - q_{p_b})^{2} + \kappa^{2} \bigr] \biggr)
\biggl[\prod_{a=1}^{N} (2\pi) \delta(q_{p_a} - q'_{p'_a}) \biggr]
\end{eqnarray}
Taking into account the constraint, Eq.(\ref{A14}), one can easily note that the only the terms
which survive in the above summation over the permutations are $P = P'$, all contributing
equal value. Thus, we finally get
\begin{equation}
 \label{A19}
\overline{\Psi^{(N)^{*}}_{\bf q'}({\bf x}) \Psi^{(N)}_{\bf q}({\bf x})} = |C|^{2} \; N! \;
\biggl(\prod_{a<b}^{N}\bigl[(q_{a} - q_{b})^{2} + \kappa^{2} \bigr] \biggr)
\biggl[\prod_{a=1}^{N} (2\pi) \delta(q_{a} - q'_{a}) \biggr]
\end{equation}
This relation defines the normalization constant
\begin{equation}
   \label{A20}
\big|C({\bf q})\big|^{2} = \frac{1}{N! \prod_{a<b}^{N} 
\bigl[ (q_{a} - q_{b})^{2} + \kappa^{2} \bigr] }
\end{equation}
 The proof of completeness of this set is given in Ref. \cite{gaudin}.
It should be noted that the above wave functions present the orthonormal set of
eigenfunctions of the problem, Eq.(\ref{A1}), for any sign of the interactions
$\kappa$, e.i. both for the repulsive, $\kappa < 0$, and for the attractive, $\kappa > 0$, cases.
However, only in the case of repulsion this set is complete, while in the 
case of attractive interactions, $\kappa > 0$, in addition to the solutions, Eq.(\ref{A11}),
which describe the continuous free particles spectrum, one finds the whole family
of discrete bound eigenstates. Detailed description of these states is given in
Appendix B.

%%%%%%%%%%%%%%%%%%%%%%%%%%%%%%%%%%%%%%%%%%%%%%%%%%%%%%%%%%%%%%%%%%%%%%%%%%%%%%%%%%%%%%%%%%%%%%%%%%%%
%%%%%%%%%%%%%%%%%%%%%%%%%%%%%%%%%%%%%%%%%%%%%%%%%%%%%%%%%%%%%%%%%%%%%%%%%%%%%%%%%%%%%%%%%%%%%%%%%%%%

\vspace{10mm}

\begin{center}

\appendix{\Large Appendix B}

\vspace{5mm}

 {\bf \large Quantum bosons with attractive interactions}

\end{center}

\newcounter{B}
\setcounter{equation}{0}
\renewcommand{\theequation}{B.\arabic{equation}}

\vspace{5mm}

{\center{\bf 1. Ground state}}

\vspace{3mm}

The simplest example of the bound eigenstate defined by eq.(\ref{A1}) (with $\kappa > 0$)
is the one in which all $N$ particles are bound into a single "cluster":
\begin{equation}
   \label{B1}
   \Psi_{q}^{(1)}({\bf x}) \; = \;  C \; 
    \exp\biggl[i q \sum_{a=1}^{N} x_{a} - \frac{1}{4}\kappa \sum_{a,b=1}^{N} |x_{a}-x_{b}| \biggr]
\end{equation}
where $C$ is the normalization constant (to be defined below) and
$q$ is the continuous momentum of free center of mass motion. Substituting
this function in Eq.(\ref{A1}), one can easily check that  this is indeed the eigenfunction
with the energy spectrum given by the relation
\begin{equation}
 \label{B2}
E \;  = \;  -\frac{1}{2\beta} \sum_{a=1}^{N} 
\biggl[iq - \frac{1}{2}\kappa \sum_{b=1}^{N} \sgn(x_{a}-x_{b}) \biggr]^{2}
\end{equation}
where it is assumed (by definition) that $\sgn(0) = 0$.
Since the result of the above summations does not depend on the mutual particles positions,
for simplicity we can order them according to Eq.(\ref{A2}). Then, using well known relations
\begin{eqnarray}
 \label{B3}
\sum_{b=1}^{N} \sgn(x_{a}-x_{b}) &=& -(N+1-2a)\\
\label{B4}
\sum_{a=1}^{N} a &=& \frac{1}{2} N (N+1)\\
\label{B5}
\sum_{a=1}^{N} a^{2} &=& \frac{N}{6} (N+1) (2N+1)
\end{eqnarray} 
for the energy spectrum, Eq.(\ref{B2}), we get
\begin{equation}
   \label{B6}
E \; = \; \frac{N}{2\beta} q^2 - \frac{\kappa^{2}}{24\beta}(N^{3}-N) \; \equiv \; 
E_{1}(q, N)
\end{equation}
The normalization constant $C$ is defined by the orthonormality
condition
\begin{equation}
   \label{B7}
\overline{\Psi_{q'}^{(1)^{*}}({\bf x}) \Psi_{q}^{(1)}({\bf x})} \; \equiv \;
\int_{-\infty}^{+\infty} dx_{1}...dx_{N} \; 
\Psi_{q'}^{(1)^{*}}({\bf x}) \Psi_{q}^{(1)}({\bf x}) \; = \; 
(2\pi) \delta(q-q')
\end{equation}
Substituting here Eq.(\ref{B1}) we get
\begin{eqnarray}
 \nonumber
\overline{\Psi_{q'}^{(1)^{*}}({\bf x}) \Psi_{q}^{(1)}({\bf x})} &=&
|C|^{2} \int_{-\infty}^{+\infty} dx_{1}...dx_{N} \; 
\exp\biggl[i (q-q') \sum_{a=1}^{N} x_{a} - 
\frac{1}{2}\kappa \sum_{a,b=1}^{N} |x_{a}-x_{b}| \biggr] 
\\
\label{B8}
&=& |C|^{2} N!
\int_{-\infty}^{+\infty} dx_{1}
\int_{x_{1}}^{+\infty} dx_{2} 
....
\int_{x_{N-1}}^{+\infty} dx_{N}
\exp\biggl[i (q-q') \sum_{a=1}^{N} x_{a} + \kappa \sum_{a=1}^{N} (N+1-2a) x_{a} \biggr]
\end{eqnarray}
where for the ordering, Eq.(\ref{A2}), we have used the relation
\begin{equation}
 \label{B9}
\frac{1}{2} \sum_{a,b=1} |x_{a}-x_{b}| \; = \; 
- \sum_{a=1}^{N} (N+1-2a) x_{a}
\end{equation}
Integrating first over $x_{N}$, then over $x_{N-1}$, and proceeding until $x_{1}$,
we find
\begin{eqnarray}
 \nonumber
\overline{\Psi_{q'}^{(1)^{*}}({\bf x}) \Psi_{q}^{(1)}({\bf x})} 
&=& |C|^{2} N! \;
\biggl(\prod_{r=1}^{N-1} \frac{1}{r[(N-r)\kappa -i(q-q')]}\biggr)
\int_{-\infty}^{+\infty} dx_{1} 
\exp\bigl[iN(q-q')x_{1}\bigr]\\
\nonumber
&=& |C|^{2} N! \; 
\biggl(\prod_{r=1}^{N-1} \frac{1}{r(N-r)\kappa}\biggr) \;
(2\pi) \delta\bigl(N(q-q')\bigr) \\
\label{B10}
&=& |C|^{2} \; \frac{N\kappa}{N!\kappa^{N}} \; (2\pi) \delta(q-q')
\end{eqnarray}
According to Eq.(\ref{B7}) this defines the normalization constant
\begin{equation}
   \label{B11}
   C \; = \; \sqrt{\frac{\kappa^N N!}{\kappa N}}  \; \equiv \;   C^{(1)}(q)
\end{equation}
Note that the eigenstate described by the considered wave function, Eq.(\ref{B1}), exists only
in the case of attraction, $\kappa > 0$, otherwise this function is divergent at infinity 
and consequently it is not normalizable.

\vspace{5mm}

It should be noted that the wave function, Eq.(\ref{B1}), can also be derived from the 
general eigenfunctions structure, Eq.(\ref{A12}), by introducing (discrete) imaginary 
parts for the momenta $q_{a}$. We assume again that the position of particles 
are ordered according to Eq.(\ref{A2}), and  define the particles momenta
according to the rule
\begin{equation}
\label{B12}
q_{a} \; = \; q - \frac{i}{2} \kappa (N+1-2a)
\end{equation}
Substituting this into Eq.(\ref{A12}) we get
\begin{eqnarray}
  \nonumber 
  \Psi_{q}^{(1)}({x_{1} < x_{2} < ... < x_{N}}) & \propto & 
\sum_{P} (-1)^{[P]} \; 
\biggl(
\prod_{a<b}^{N}
\biggl[ 
  \Bigl( q - \frac{i}{2} \kappa (N+1-2P_{a})\Bigr) 
- \Bigl( q - \frac{i}{2} \kappa (N+1-2P_{b})\Bigr) -i \kappa 
\biggr]
\biggr) \times
\\
\nonumber
&&\times
\exp\biggl[ iq\sum_{a=1}^{N}x_{a} +\frac{\kappa}{2}\sum_{a=1}^{N} (N+1-2P_{a}) x_{a} \biggr]  
\\
\nonumber
\\
& \propto &
\sum_{P} (-1)^{[P]} 
\biggl(\prod_{a<b}^{N}\bigl[ P_{b}-P_{a} + 1 \bigr] \biggr) \;
\exp\biggl[ iq\sum_{a=1}^{N}x_{a} +\frac{\kappa}{2}\sum_{a=1}^{N} (N+1-2P_{a}) x_{a} \biggr] 
\label{B13}
\end{eqnarray}
Here one can easily note that due to the presence of the product 
$\prod_{a<b}^{N}[P_{b}-P_{a} + 1] $ in the summation over permutations only the 
trivial one, $P_{a} = a$, gives non-zero contribution (if we permute
any two numbers in the sequence $1, 2, ... , N$ then we can always find two numbers
$a < b$, such that $P_{b} = P_{a} - 1$). Thus
\begin{equation}
 \label{B14}
 \Psi_{q}^{(1)}({x_{1} < x_{2} < ... < x_{N}}) \; \propto \;
\exp\biggl[ iq\sum_{a=1}^{N}x_{a} +\frac{\kappa}{2}\sum_{a=1}^{N} (N+1-2a) x_{a} \biggr] 
\end{equation}
Taking into account the relation, Eq.(\ref{B9}), we recover the function, Eq.(\ref{B1}),
which is symmetric with respect to its $N$ arguments and therefore can be extended
beyond the sector, Eq.(\ref{A2}), for arbitrary particles positions.
Finally, substituting the momenta, Eq.(\ref{B12}), into the general expression for the energy 
spectrum, Eq.(\ref{A13}), we get
\begin{equation}
 \label{B15}
E \, = \; \frac{1}{2\beta} \sum_{a=1}^{N} \bigl[q - \frac{i}{2} \kappa (N+1-2a) \bigr]^{2} 
\end{equation}
Performing here simple summations (using Eqs.(\ref{B4}), (\ref{B5})) 
one recovers Eq.(\ref{B6}).

\vspace{10mm}

{\center{\bf 2. Eigenfunctions}}

\vspace{5mm}

A generic eigenfunction of attractive bosons is characterized by $N$ momenta parameters 
$\{ q_{a}\} \; (a= 1,2,...N)$ which  may have imaginary parts.
It is convenient to group these parameters into $M \; \; (1 \leq M \leq N)$ "vector" momenta, 
\begin{equation}
\label{B16}
q^{\alpha}_{r} \; = \; 
q_{\alpha} \; - \; \frac{i}{2} \; \kappa \; (n_{\alpha} + 1 -2 r)
\end{equation}
where $q_{\alpha} \; (\alpha = 1, 2, ..., M)$ are the continuous (real) parameters,
and the discrete imaginary components of each "vector"  are labeled by an index 
$r = 1, 2, ..., n_{\alpha}$. With the given total number of particles equal to
$N$, the integers $n_\alpha$ have to satisfy the constraint
\begin{equation}
\label{B17}
\sum_{\alpha = 1}^{M} \; n_{\alpha} \; = \; N
\end{equation}
In other words, a generic eigenstate is characterized by the discrete number $M$
of complex ''vector`` momenta, by the set
of $M$ integer parameters $\{ n_{1}, n_{2}, ..., n_{M} \} \equiv {\bf n}$ (which are the numbers
of imaginary components of each ''vector``) and by the set of $M$ real continuous
momenta $\{ q_{1}, q_{2}, ..., q_{M} \}  \equiv {\bf q}$. 

The general expression for the eigenfunctions is given in Eqs.(\ref{A9})-(\ref{A12}).
To understand the structure of the determinant of the $N\times N$ matrix
$\exp( i  q_{a}  x_{b})$, which defines these wave functions, 
the $N$ momenta $q_{a}$, eq.(\ref{B16}), can be ordered as follows:
\begin{equation}
 \label{B18}
\{ q_{a} \} \; \equiv \; \{q^{\alpha}_{r}\} \; = \; 
                    \{ q^{1}_{1}, \; q^{1}_{2}, \; ... \; , \; q^{1}_{n_{1}}; \; 
                       q^{2}_{1}, \; q^{2}_{2}, \; ... \; , \; q^{2}_{n_{2}}; \; ... \; ; \; 
                       q^{M}_{1}, \; q^{M}_{2}, \; ... \; , \; q^{M}_{n_{M}} \}
\end{equation}
By definition,
\begin{equation}
\label{B19}
    \det\Bigl[\exp( i q_a \, x_c)\Bigr]_{(c,d)=1,...,N} \; = \; 
\sum_{P} (-1)^{[P]} \; \exp\Bigl[i \sum_{a=1}^{N} q_{p_{a}} x_{a} \Bigr]
\end{equation}
where the summation goes over the permutations of $N$ momenta $\{ q_{a}\}$, Eq.(\ref{B18}),
over $N$ particles $\{ x_{1}, x_{2}, ..., x_{N}\}$, and $[P]$ denotes the parity of the permutation.
For a given permutation $P$ a particle number $a$ 
is attributed a momentum component $q^{\alpha(a)}_{r(a)}$.
The particles getting the momenta with the same 
 $\alpha$ (having the same real part $q_{\alpha}$) will be called  
belonging to a cluster $\Omega_{\alpha}$. For a given permutation $P$
the particles belonging 
to the same cluster are  numbered by the  "internal" index $r = 1,...,n_{\alpha}$.
Thus, according to Eq.(\ref{A11}),
\begin{equation}
 \label{B20}
\Psi_{\bf q, n}^{(M)}({\bf x}) \; = \;  C^{(M)}_{\bf q, n}
\sum_{P} (-1)^{[P]} 
    \biggl(\prod_{a<b}^{N}\Bigl[ -i\bigl(\partial_{x_a} - \partial_{x_b}\bigr) +i \kappa \sgn(x_{a}-x_{b})\Bigr]\biggr) \;
    \exp\biggl[ i \sum_{c=1}^{N} q^{\alpha(c)}_{r(c)} \, x_{c} \biggr]
\end{equation}
where $C^{(M)}_{\bf q, n}$ is the normalization constant to be defined later.
Substituting here Eq.(\ref{B16}) and taking derivatives we get
\begin{eqnarray}
   \label{B21}
\nonumber
  \Psi_{\bf q, n}^{(M)}({\bf x}) &=& C^{(M)}_{\bf q, n}
\sum_{P} (-1)^{[P]} 
\prod_{a<b}^{N}
\biggl[ 
  \Bigl(q_{\alpha(a)} - \frac{i\kappa}{2} \big[n_{\alpha(a)} + 1 - 2r(a)\big] \Bigr)
- \Bigl(q_{\alpha(b)} - \frac{i\kappa}{2} \big[n_{\alpha(b)} + 1 - 2r(b)\big] \Bigr) 
+i\kappa \sgn(x_{a}-x_{b})
\biggr]  
\times\\
& &\times\exp\biggl[
i\sum_{c=1}^{N}q_{\alpha(c)} x_{c} 
 +\frac{\kappa}{2}\sum_{c=1}^{N} \Bigl(n_{\alpha(c)} +1 - 2r(c)\Bigr) x_{c} \biggr]
\end{eqnarray}
The pre-exponential product in the above equation contains two types of term: the pairs of 
points $(a,b)$ which belong to different clusters ($\alpha(a)\not=\alpha(b)$), and pairs of points
which belong to the same cluster ($\alpha(a) = \alpha(b)$). In the last case, the product
$\Pi_{\alpha}$ over the pairs of points which belong to a cluster $\Omega_{\alpha}$ 
reduces to 
\begin{equation}
 \label{B22}
\Pi_{\alpha} \; \propto \; 
\prod_{a<b\in\Omega_{\alpha}}\bigl[r(b) - r(a) -\sgn(x_{a}-x_{b})]\bigr] 
\end{equation}
Similarly to the ground state wave function Eq.\ (\ref{B13})--(\ref{B14}),
one can easily note that due to the presence of this product 
in the summations over $n_{\alpha}!$ ''internal`` (inside the cluster $\Omega_{\alpha}$)
permutations $r(a)$ 
only one permutation gives non-zero contribution. 
To prove this statement, we note
that the wave function $\Psi_{\bf q, n}^{(M)}({\bf x})$ is symmetric with
respect to permutations of its $N$ arguments $\{x_{a}\}$; it is then
sufficient to consider the case where the positions of the particles are
ordered, $x_1 < x_2 < \dots < x_N$. In particular,
the particles $\{x_{a_{k}}\} \; (k = 1, 2, \dots, n_{\alpha})$
belonging to the same cluster $\Omega_{\alpha}$ are also ordered $x_{a_{1}} <
x_{a_{2}} < \dots < x_{a_{n_{\alpha}}}$.  In this case
\begin{equation}
 \label{B23}
\Pi_{\alpha} \; \propto \; 
\prod_{k<l}^{n_{\alpha}} \bigl[r(l) - r(k) + 1\bigr] 
\end{equation}
Now it is evident that the above product is non-zero only for the trivial
permutation, $r(k) = k$ (since if we permute
any two numbers in the sequence $1, 2, ... , n_{\alpha}$, we can always find two numbers
$k < l$, such that $r(l) = r(k) - 1$). In this case
\begin{equation}
 \label{B24}
\Pi_{\alpha} \; \propto \; 
\prod_{k<l}^{n_{\alpha}} \bigl[ l - k + 1 \bigr] 
\end{equation}
Including the values of all these "internal" products, 
Eq.(\ref{B24}), into the redefined normalization constant
$C^{(M)}_{\bf q, n}$, for the wave function, Eq.(\ref{B21}) 
(with  $x_1 < x_2 < \dots < x_N$), we obtain
\begin{eqnarray}
   \label{B25}
\nonumber
  \Psi_{\bf q, n}^{(M)}({\bf x}) &=& C^{(M)}_{\bf q, n}
{\sum_{P}}' (-1)^{[P]} 
\prod_{\substack{ a<b\\ \alpha(a)\not=\alpha(b) }}^{N}  
\biggl[ 
\Bigl(q_{\alpha(a)} - \frac{i\kappa}{2} n_{\alpha(a)}\Bigr) -
\Bigl(q_{\alpha(b)} - \frac{i\kappa}{2} n_{\alpha(b)}\Bigr) 
+i\kappa\Bigl(r(a) - r(b) - 1)\Bigr)\Biggr]  
\times\\
& &\times\exp\biggl[
i\sum_{c=1}^{N}q_{\alpha(c)} x_{c} 
 +\frac{\kappa}{2}\sum_{c=1}^{N} \Bigl(n_{\alpha(c)} +1 - 2r(c)\Bigr) x_{c} \biggr]
\end{eqnarray}
where  the product  goes only 
 over the pairs of particles belonging to {\it different}
clusters, and the symbol ${\sum_{P}}'$ means 
that the summation goes only over 
the permutations $P$ in which the "internal" indexes
$r(a)$ are ordered inside each cluster.

Now taking into account the symmetry of the wave function $\Psi_{\bf q, n}^{(M)}({\bf x})$
with respect to the permutations of its arguments the expression in Eq.(\ref{B25}) 
can be easily continued beyond the
the sector $x_1 < x_2 < ... < x_N$ for the entire coordinate space $R_{N}$.
Using the relations
\begin{equation}
 \label{B26}
\sum_{a\in\Omega_{\alpha}} \bigl(n_{\alpha} +1 - 2r(a)\bigr) x_{a} \; = \; 
\sum_{k=1}^{n_{\alpha}} (n_{\alpha} +1 - 2 k) x_{a_{k}} \; = \; 
-\frac{1}{2} \sum_{k,l=1}^{n_{\alpha}}
\big|x_{a_{k}} - x_{a_{l}}\big|
\end{equation}
(where $x_{a_{1}} < x_{a_{2}} < ... < x_{a_{n_{\alpha}}}$), 
for the wave function $\Psi_{\bf q, n}^{(M)}({\bf x})$ with {\it arbitrary} particles positions 
we get the following sufficiently compact representation (cf. Eq.(\ref{B20})):
\begin{equation}
 \label{B27}
 \Psi_{\bf q, n}^{(M)}({\bf x}) =  C^{(M)}_{\bf q, n}
{\sum_{P}}' (-1)^{[P]} 
\prod_{\substack{ a<b\\ \alpha(a)\not=\alpha(b) }}^{N} 
\biggl[ -i\bigl(\partial_{x_a} - \partial_{x_b}\bigr) 
+ i \kappa \sgn(x_{a}-x_{b})\biggr]
 \exp\biggl[
i\sum_{\alpha=1}^{M}q_{\alpha}\sum_{c\in\Omega_{\alpha}}^{n_{\alpha}} x_{c} 
 -\frac{\kappa}{4}\sum_{\alpha=1}^{M}\sum_{c,c'\in\Omega_{\alpha}}^{n_{\alpha}} |x_{c}-x_{c'}| \biggr]
\end{equation}
Note that although the positions of particles belonging to the same cluster are 
ordered, the mutual positions of particles belonging to different clusters could be arbitrary,
so that geometrically the clusters are free to "penetrate" each other. In other words, 
the name "cluster" does not assume geometrically compact particles positions.

Finally, substituting Eq.(\ref{B16})-(\ref{B17}) into Eq.(\ref{A13}), 
for the energy spectrum one easily obtains:
\begin{equation}
\label{B28}
E_{M}({\bf q,n}) \; = \;
\frac{1}{2\beta} \sum_{\alpha=1}^{M} \; \sum_{r=1}^{n_{\alpha}} (q^{\alpha}_{k})^{2} 
\; = \; \frac{1}{2\beta} \sum_{\alpha=1}^{M} \; n_{\alpha} q_{\alpha}^{2} \;
- \; \frac{\kappa^{2}}{24\beta}\sum_{\alpha=1}^{M} (n_{\alpha}^{3}-n_{\alpha})
\end{equation}

\vspace{10mm}

{\center {\bf 3. Orthonormality}}

\vspace{3mm}

We define the overlap of two wave functions characterized by two sets of
parameters, $(M, {\bf n}, {\bf q})$ and $(M', {\bf n'}, {\bf q'})$ as
\begin{equation}
\label{B29}
Q^{(M,M')}_{{\bf n},{\bf n'}}({\bf q}, {\bf q'}) \; \equiv \; 
\int_{-\infty}^{+\infty} d^{N}{\bf x} \; 
 \Psi_{\bf q', n'}^{(M')^{*}}({\bf x})
 \Psi_{\bf q, n}^{(M)}({\bf x})  
\end{equation}
Substituting here Eq.(\ref{B27}) we get
\begin{eqnarray}
\label{B30}
Q^{(M,M')}_{{\bf n},{\bf n'}}({\bf q}, {\bf q'}) &=&
C^{(M)}_{\bf q, n} C^{(M')^{*}}_{\bf q', n'} 
{\sum_{P}}' {\sum_{P'}}' (-1)^{[P] + [P']}  \; 
\int_{-\infty}^{+\infty} d^{N}{\bf x}  
\\
\nonumber
\\
\nonumber
&& \times
\biggl(
\prod_{\substack{ a<b\\ \alpha'(a)\not=\alpha'(b) }}^{N} \biggl[ 
i \bigl(\partial_{x_a} - \partial_{x_b}\bigr) - i \kappa \sgn(x_{a}-x_{b}) \biggr] \biggr)
\exp\biggl[
-i\sum_{\alpha=1}^{M'}q'_{\alpha}\sum_{c\in\Omega'_{\alpha}}^{n'_{\alpha}} x_{c} 
 -\frac{\kappa}{4}\sum_{\alpha=1}^{M'}\sum_{c,c'\in\Omega'_{\alpha}}^{n'_{\alpha}} |x_{c}-x_{c'}| \biggr]
\times
\\
\nonumber
\\
\nonumber
&& \times
\biggl(
\prod_{\substack{ a<b\\ \alpha(a)\not=\alpha(b) }}^{N} \biggl[ 
-i \bigl(\partial_{x_a} - \partial_{x_b}\bigr) +i \kappa \sgn(x_{a}-x_{b}) \biggr] \biggr)
\exp\biggl[
i\sum_{\alpha=1}^{M}q_{\alpha}\sum_{c\in\Omega_{\alpha}}^{n_{\alpha}} x_{c} 
 -\frac{\kappa}{4}\sum_{\alpha=1}^{M}\sum_{c,c'\in\Omega_{\alpha}}^{n_{\alpha}} |x_{c}-x_{c'}| \biggr]
\end{eqnarray}
where $\{\Omega_{\alpha}\}$ and $\{\Omega'_{\alpha}\}$ denote the clusters of the permutations $P$
and $P'$ correspondingly. Integrating by parts we obtain
\begin{eqnarray}
\nonumber
Q^{(M,M')}_{{\bf n},{\bf n'}}({\bf q}, {\bf q'}) &=&
C^{(M)}_{\bf q, n} C^{(M')^{*}}_{\bf q', n'} 
{\sum_{P}}' {\sum_{P'}}' (-1)^{[P] + [P']}  
\int_{-\infty}^{+\infty} d^{N}{\bf x}  \; 
\exp\biggl[
-i\sum_{\alpha=1}^{M'}q'_{\alpha}\sum_{c\in\Omega'_{\alpha}}^{n'_{\alpha}} x_{c} 
 -\frac{\kappa}{4}\sum_{\alpha=1}^{M'}\sum_{c,c'\in\Omega'_{\alpha}}^{n'_{\alpha}} |x_{c}-x_{c'}| \biggr] \times
\\
\nonumber
\\
\nonumber
&& \times
\biggl(
\prod_{\substack{ a<b\\ \alpha'(a)\not=\alpha'(b) }}^{N} \biggl[ 
-i\bigl(\partial_{x_a} - \partial_{x_b}\bigr) - i \kappa \sgn(x_{a}-x_{b}) \biggr] \biggr)
\biggl(
\prod_{\substack{ a<b\\ \alpha(a)\not=\alpha(b) }}^{N} \biggl[ 
-i\bigl(\partial_{x_a} - \partial_{x_b}\bigr) +i \kappa \sgn(x_{a}-x_{b}) \biggr] \biggr) \times
\\
\nonumber
\\
&& \times
\exp\biggl[
i\sum_{\alpha=1}^{M}q_{\alpha}\sum_{c\in\Omega_{\alpha}}^{n_{\alpha}} x_{c} 
 -\frac{\kappa}{4}\sum_{\alpha=1}^{M}\sum_{c,c'\in\Omega_{\alpha}}^{n_{\alpha}} |x_{c}-x_{c'}| \biggr]
\label{B31}
\end{eqnarray}

First, let us consider the case when the integer parameters of the two functions
coincide, $M = M'$, $\; {\bf n} = {\bf n'}$, and for the moment let us suppose that 
all these integer parameters $\{n_{\alpha}\}$ are {\it different},
$1 \leq n_{1} < n_{2} < ... < n_{M}$. Then, in the summations over the permutations
in Eq.(\ref{B31}), we find two types of terms:

(A) the "diagonal" ones in which the two
permutations coincide, $P = P'$ ;

(B) the "off-diagonal" ones in which
the two permutations are different, $P \not= P'$.

The contribution of the "diagonal" ones reeds
\begin{eqnarray}
\nonumber
Q^{(M,M)^{(A)}}_{{\bf n},{\bf n}}({\bf q}, {\bf q'}) &=&
C^{(M)}_{\bf q, n} C^{(M)^{*}}_{\bf q', n} \;
{\sum_{P}}'
\int_{-\infty}^{+\infty} d^{N}{\bf x}  \; 
\exp\biggl[
-i\sum_{\alpha=1}^{M} q'_{\alpha}\sum_{c\in\Omega_{\alpha}}^{n_{\alpha}} x_{c} 
 -\frac{\kappa}{4}\sum_{\alpha=1}^{M}\sum_{c,c'\in\Omega_{\alpha}}^{n_{\alpha}} |x_{c}-x_{c'}| \biggr] \times
\\
\nonumber
\\
&& \times
\biggl(
\prod_{\substack{ a<b\\ \alpha(a)\not=\alpha(b) }}^{N} \biggl[ 
-\bigl(\partial_{x_a} - \partial_{x_b}\bigr)^{2} + \kappa^{2} \biggr] \biggr) \;
\exp\biggl[
i\sum_{\alpha=1}^{M}q_{\alpha}\sum_{c\in\Omega_{\alpha}}^{n_{\alpha}} x_{c} 
 -\frac{\kappa}{4}\sum_{\alpha=1}^{M}\sum_{c,c'\in\Omega_{\alpha}}^{n_{\alpha}} 
|x_{c}-x_{c'}| \biggr]
\label{B32}
\end{eqnarray}
It is evident that all permutations $\alpha(a)$ in the above equation give the same contribution
and therefore it is sufficient to consider only the contribution of the "trivial"
permutation which is represented by  Eq.(\ref{B18}). The cluster  ordering
given by this permutation we denote by $\alpha_{0}(a)$. For this particular configuration
of clusters we can redefine the particles numbering, so that instead of a "plane"
index $a = 1, 2, ..., N$ the particles would be counted by two indexes $(\alpha, r)$:
$\{x_{a}\} \to \{x_{r}^{\alpha}\} \; (\alpha = 1, ..., M) \; (r = 1, ..., n_{\alpha})$
indicating to which cluster $\alpha$ a given particle belongs and what is
its "internal" cluster number $r$. Due to the symmetry of the integrated expression
in Eq.(\ref{B32}) with respect to the permutations of the particles inside the clusters,
we can introduce the "internal" particles ordering 
for every cluster: $x_{1}^{\alpha}  < x_{2}^{\alpha} < ... < x_{n_{\alpha}}^{\alpha}$. 
In this way, using the relation, Eq.(\ref{B26}), we get
\begin{eqnarray}
\nonumber
&& Q^{(M,M)^{(A)}}_{{\bf n},{\bf n}}({\bf q}, {\bf q'}) \; = \; 
C^{(M)}_{\bf q, n} C^{(M)^{*}}_{\bf q', n} \; \frac{N!}{n_{1}! n_{2}! ... n_{M}!}
\biggl[\prod_{\alpha=1}^{M}
\biggl(
n_{\alpha}!
\int_{-\infty}^{+\infty} dx_{1}^{\alpha}
\int_{x_{1}^{\alpha}}^{+\infty} dx_{2}^{\alpha}
....
\int_{x_{n_{\alpha}-1}^{\alpha}}^{+\infty} dx_{n_{\alpha}}^{\alpha}
\biggr) \biggr] \times
\\
\nonumber
\\
\nonumber
&& \times
\exp\biggl[
-i\sum_{\alpha=1}^{M} q'_{\alpha}\sum_{r=1}^{n_{\alpha}} x_{r}^{\alpha} 
+\frac{\kappa}{2}\sum_{\alpha=1}^{M}\sum_{r=1}^{n_{\alpha}} (n_{\alpha} +1 - 2r)x_{r}^{\alpha} \biggr] \times
\\
\nonumber
\\
&& \times
\biggl(
\prod_{\alpha < \beta}^{M} \prod_{r=1}^{n_{\alpha}} \prod_{r'=1}^{n_{\beta}}
\biggl[ 
-\bigl(\partial_{x_{r}^{\alpha}} - \partial_{x_{r'}^{\beta}}\bigr)^{2} + \kappa^{2} \biggr] \biggr) 
\exp\biggl[
i\sum_{\alpha=1}^{M} q_{\alpha}\sum_{r=1}^{n_{\alpha}} x_{r}^{\alpha} 
+\frac{\kappa}{2}\sum_{\alpha=1}^{M}\sum_{r=1}^{n_{\alpha}} (n_{\alpha} +1 - 2r)x_{r}^{\alpha}\biggr]
\label{B33}
\end{eqnarray}
where the factor $N!/n_{1}!...n_{M}!$ is the total number of permutations of $M$ clusters
over $N$ particles. Taking the derivatives and reorganizing the terms we obtain
\begin{eqnarray}
\nonumber
Q^{(M,M)^{(A)}}_{{\bf n},{\bf n}}({\bf q}, {\bf q'}) &=&
C^{(M)}_{\bf q, n} C^{(M)^{*}}_{\bf q', n} \; N! \; 
\biggl(
\prod_{\alpha < \beta}^{M} \prod_{r=1}^{n_{\alpha}} \prod_{r'=1}^{n_{\beta}}
\bigg| 
\bigl(q_{\alpha} - \frac{i\kappa}{2} n_{\alpha}\bigr) -
\bigl(q_{\beta}  - \frac{i\kappa}{2} n_{\beta} \bigr) +
i\kappa \; (r - r' - 1) \bigg|^{2} \biggr) \times
\\
\nonumber
\\
&& \times
\prod_{\alpha=1}^{M} \Biggr\{
\int_{-\infty}^{+\infty} dx_{1}^{\alpha}
\int_{x_{1}^{\alpha}}^{+\infty} dx_{2}^{\alpha}
....
\int_{x_{n_{\alpha}-1}^{\alpha}}^{+\infty} dx_{n_{\alpha}}^{\alpha} \; 
\mbox{\Large e}^{ \;
i(q_{\alpha}-q'_{\alpha}) \sum_{r=1}^{n_{\alpha}} x_{r}^{\alpha} 
+\kappa\sum_{r=1}^{n_{\alpha}} (n_{\alpha} +1 - 2r)x_{r}^{\alpha}}
\Biggr\}
\label{B34}
\end{eqnarray}
Simple integrations over $x_{r}^{\alpha}$ yields (cf. Eqs.(\ref{B8})-(\ref{B10}))
\begin{eqnarray}
\nonumber
Q^{(M,M)^{(A)}}_{{\bf n},{\bf n}}({\bf q}, {\bf q'}) &=&
\big|C^{(M)}_{\bf q, n}\bigr|^{2}  \; N! \; 
\biggl(
\prod_{\alpha < \beta}^{M} \prod_{r=1}^{n_{\alpha}} \prod_{r'=1}^{n_{\beta}}
\bigg| 
\bigl(q_{\alpha} - \frac{i\kappa}{2} n_{\alpha}\bigr) -
\bigl(q_{\beta}  - \frac{i\kappa}{2} n_{\beta} \bigr) +
i\kappa \; (r - r' - 1) \bigg|^{2} \biggr) \times
\\
\nonumber
\\
&& \times
\prod_{\alpha=1}^{M} \Biggr[
\frac{n_{\alpha}\kappa}{(n_{\alpha}!)^{2} \kappa^{n_{\alpha}}} \; 
(2\pi) \delta(q_{\alpha} - q'_{\alpha}) 
\Biggr]
\label{B35}
\end{eqnarray}

Now let us prove that the  "off-diagonal" terms of Eq.(\ref{B31}),
in which the permutations $P$ and $P'$ are different, give no contribution.
Here we can also chose one of the permutations, 
say the permutation $P$, to be the "trivial" one 
represented by  Eq.(\ref{B18}) with the cluster  ordering
denoted by $\alpha_{0}(a)$. Given the symmetry of the wave functions 
it will be sufficient to consider the contribution of the sector
$x_{1} < x_{2} < ... < x_{N}$.
According to Eq.(\ref{B31}), we get
\begin{eqnarray}
\nonumber
Q^{(M,M)^{(B)}}_{{\bf n},{\bf n}}({\bf q}, {\bf q'}) &\propto&
 {\sum_{P'}}' (-1)^{[P']}  
\int_{x_{1} < ... < x_{N}} d^{N}{\bf x}  \; 
\exp\biggl[
-i\sum_{\alpha=1}^{M}q'_{\alpha}\sum_{a\in\Omega'_{\alpha}}^{n_{\alpha}} x_{a} 
 -\frac{\kappa}{4}\sum_{\alpha=1}^{M}\sum_{a,b\in\Omega'_{\alpha}}^{n_{\alpha}} |x_{a}-x_{b}| \biggr] \times
\\
\nonumber
\\
\nonumber
&& \times
\biggl(
\prod_{\substack{ a<b\\ \alpha_{0}(a)\not=\alpha_{0}(b) }}^{N} \biggl[ 
-i\bigl(\partial_{x_a} - \partial_{x_b}\bigr) +i \kappa \sgn(x_{a}-x_{b}) \biggr] \biggr)
\biggl(
\prod_{\substack{ a<b\\ \alpha'(a)\not=\alpha'(b) }}^{N} \biggl[ 
-i\bigl(\partial_{x_a} - \partial_{x_b}\bigr) -i \kappa \sgn(x_{a}-x_{b}) \biggr] \biggr)
\\
\nonumber
\\
&& \times
\exp\biggl[
i\sum_{\alpha=1}^{M}q_{\alpha}\sum_{a\in\Omega^{o}_{\alpha}}^{n_{\alpha}} x_{a} 
 +\frac{\kappa}{2}\sum_{\alpha=1}^{M}\sum_{a\in\Omega^{o}_{\alpha}}^{n_{\alpha}} (n_{\alpha}+1-2r(a))x_{a} \biggr]
\label{B36}
\end{eqnarray}
Here the symbols $\{\Omega^{o}_{\alpha}\}$  denote the clusters 
of the trivial permutation $\alpha_{0}(a)$.
Since $P' \not= P$,
some of the clusters $\Omega'_{\alpha}$ must be different from $\Omega^{o}_{\alpha}$.
As an illustration, let us consider a particular case of $N=10$, with three clusters
$n_{1}=5$ (denoted by the symbol "$\bigcirc$") , $n_{2}=2$ (denoted by the symbol "$\times$")
and $n_{3}=3$ (denoted by the symbol "$\triangle$"):

\vspace{5mm}

\begin{center}

\begin{tabular}{|c||c|c|c|c|c|c|c|c|c|c|}
 \hline
particle number $a$ & 1 & 2 & 3 & 4 & 5 & 6 & 7 & 8 & 9 & 10 \\
\hline
permutation $\alpha_{0}(a)$ &$\bigcirc$&$\bigcirc$&$\bigcirc$&$\bigcirc$&$\bigcirc$&$\times$&$\times$&$\triangle$&$\triangle$&$\triangle$ \\
\hline
permutation $\alpha'(a)$ &$\bigcirc$&$\bigcirc$&$\bigcirc$&$\triangle$&$\bigcirc$&$\times$&$\times$&$\bigcirc$&$\triangle$&$\triangle$ \\
\hline
\end{tabular}

\end{center}

\vspace{5mm}
Here in the permutation $\alpha'(a)$ the particle $a=4$ belong to the cluster $\alpha=3$ 
(and not to the cluster $\alpha=1$ as in the permutation $\alpha_{0}(a)$), and the 
particle $a=8$ belong to the cluster $\alpha=1$ (and not to the cluster $\alpha=3$
as in the permutation $\alpha_{0}(a)$). Now let us look carefully at the structure of the 
products in Eq.(\ref{B36}). Unlike the first product, which contains no "internal"
products among particles belonging to the cluster $\Omega^{o}_{1}$, the second product
{\it does}. Besides, the signs of the differential operators
$\bigl(\partial_{x_a} - \partial_{x_b}\bigr)$ in the second product 
is {\it opposite} to the "normal" ones in the first product (cf. Eqs.(\ref{B22})-(\ref{B24})).
It is these two factors (the presence of the "internal" products and the "wrong" signs
of the differential operators) which makes the "off-diagonal" contributions, Eq.(\ref{B36}),
to be zero. Indeed, in the above example, the second product contains the term
\begin{equation}
 \label{B37}
\Pi'_{4,5} \; \equiv \; 
\biggl[-i\bigl(\partial_{x_4} - \partial_{x_5}\bigr) + i \kappa  \biggr] 
\exp\biggl[
i\sum_{\alpha=1}^{3} q_{\alpha}\sum_{a\in\Omega^{o}_{\alpha}}^{n_{\alpha}} x_{a} 
 +\frac{\kappa}{2}\sum_{\alpha=1}^{3}\sum_{a\in\Omega^{o}_{\alpha}}^{n_{\alpha}} (n_{\alpha}+1-2r(a))x_{a} \biggr]
\end{equation}
(we remind that the particles in the clusters $\Omega^{o}_{\alpha}$ are ordered,
and in particular  $x_{4} < x_{5}$).
Taking the derivatives, we get
\begin{eqnarray}
 \nonumber
\Pi'_{4,5}  &\propto&  
\biggl[-\biggl(iq_{1} + \frac{\kappa}{2}\bigl(n_{1}+1-2r(4)\bigr) 
             - iq_{1} - \frac{\kappa}{2}\bigl(n_{1}+1-2r(5)\bigr) \biggr) + \kappa \Biggr] \\
&\propto&
\bigl[ r(4)-r(5) + 1 \bigr] \; = \; 0
\label{B38}
\end{eqnarray}
since in the first cluster $r(a)=a$. 

One can easily understand that the above example reflect 
the general situation. Since all the cluster sizes $n_{\alpha}$ are supposed to be different,
whatever the permutation $\alpha'(a)$ is, we can always find a cluster
$\Omega^{o}_{\alpha}$ 
such that some of its particles belong to the same cluster number $\alpha$ in the permutation
$\alpha'(a)$ while the others do not. Then one has to consider the contribution of
the product of two {\it neighboring number} points
\begin{equation}
 \label{B39}
\Pi'_{k, k+1} \; = \; 
\biggl[-i\bigl(\partial_{x_k} - \partial_{x_{k+1}}\bigr) + i \kappa  \biggr] 
\exp\biggl[
i\sum_{\alpha=1}^{M} q_{\alpha}\sum_{a\in\Omega^{o}_{\alpha}}^{n_{\alpha}} x_{a} 
 +\frac{\kappa}{2}\sum_{\alpha=1}^{M}\sum_{a\in\Omega^{o}_{\alpha}}^{n_{\alpha}} (n_{\alpha}+1-2r(a))x_{a} \biggr]
\end{equation}
where in the permutation $\alpha'(a)$ the particle $k$ belong to the cluster number $\alpha$ 
and the particle $(k+1)$ belong to some
other cluster. Taking the derivatives
one gets
\begin{equation}
 \label{B40}
\Pi'_{k, k+1} \; \propto \; 
\bigl[r(k)-r(k+1) + 1\bigr] \; = \; 0
\end{equation}
as $r(a)$ is the ''internal`` particle number in the cluster $\Omega^{o}_{\alpha}$,
where $r(k+1) = r(k) + 1$ (cf. Eqs.(\ref{B22})-(\ref{B24})).

Thus, the only non-zero contribution to the overlap, Eq.(\ref{B29}), of two wave function 
$\Psi_{\bf q', n}^{(M)}({\bf x})$ and $\Psi_{\bf q, n}^{(M)}({\bf x})$ (having the same
number of clusters $M$ and characterized by
the same set of the integer parameters  $1 \leq n_{1} < n_{2} < ... < n_{M}$) comes from the "diagonal" terms, Eq.(\ref{B35}):
\begin{eqnarray}
\nonumber
Q^{(M,M)}_{{\bf n},{\bf n}}({\bf q}, {\bf q'}) &=&
\big|C^{(M)}_{\bf q, n}\big|^{2} \; N!  
\prod_{\alpha=1}^{M} \Biggr[
\frac{n_{\alpha}\kappa}{(n_{\alpha}!)^{2} \kappa^{n_{\alpha}}} 
\Biggr] 
\biggl(
\prod_{\alpha < \beta}^{M} \prod_{r=1}^{n_{\alpha}} \prod_{r'=1}^{n_{\beta}}
\bigg| 
\bigl(q_{\alpha} - \frac{i\kappa}{2} n_{\alpha} \bigr) -
\bigl(q_{\beta}  - \frac{i\kappa}{2} n_{\beta}  \bigr)
+i\kappa \; (r - r' - 1)
\bigg|^{2} \biggr) 
\times
\\
\nonumber
\\
&& \times
\prod_{\alpha=1}^{M} \Biggr[
(2\pi) \delta(q_{\alpha} - q'_{\alpha}) 
\Biggr]
\label{B41}
\end{eqnarray}

The situation when there are clusters which have the same numbers of particles $n_{\alpha}$ 
is somewhat more complicated. Let us consider the overlap between two wave function
$\Psi_{\bf q', n}^{(M)}({\bf x})$ and $\Psi_{\bf q, n}^{(M)}({\bf x})$ (which, as before
have the same $M$ and ${\bf n}$) such that in the set of $M$ integers
$n_{1}, n_{2}, ... , n_{M}$ there are two $n_{\alpha}$'s which are equal, say
$n_{\alpha_{1}} = n_{\alpha_{2}}$ (where $\alpha_{1} \not= \alpha_{2}$).
In the eigenstate $({\bf q', n})$ these two clusters have 
the center of mass momenta $q'_{\alpha_{1}}$ and $q'_{\alpha_{2}}$, and in the 
the eigenstate $({\bf q, n})$ they have the momenta $q_{\alpha_{1}}$ and $q_{\alpha_{2}}$
correspondingly.
According to the above discussion, the non-zero contributions in the summation
over the cluster permutations $\alpha(a)$ and $\alpha'(a)$ in Eq.(\ref{B31})
appears only if the clusters $\{\Omega_{\alpha}\}$ of the permutation $\alpha(a)$
totally coincide with the clusters $\{\Omega'_{\alpha}\}$ of the permutation $\alpha'(a)$.
In the case when all $n_{\alpha}$ are different this is possible only if 
the permutation $\alpha(a)$ coincides with the permutation $\alpha'(a)$. In contrast to that,
in the case when we have $n_{\alpha_{1}} = n_{\alpha_{2}}$, there are {\it two}
non-zero options. The first one, as before, is given by the "diagonal" terms
with $\alpha(a) = \alpha'(a)$ (so that the clusters $\{\Omega_{\alpha}\}$ and
$\{\Omega'_{\alpha}\}$ are just the same), and this contribution
is proportional to $\delta(q_{\alpha_{1}} - q'_{\alpha_{1}}) \, \delta(q_{\alpha_{2}} - q'_{\alpha_{2}})$.
The second  ("off-diagonal") contribution is given by such permutation $\alpha'(a)$
in which the cluster $\Omega'_{\alpha_{1}}$ (of the permutation $\alpha'(a)$) coincide
with the cluster $\Omega_{\alpha_{2}}$ (of the permutation $\alpha(a)$) and 
the cluster $\Omega'_{\alpha_{2}}$ (of the permutation $\alpha'(a)$) coincide
with the cluster $\Omega_{\alpha_{1}}$ (of the permutation $\alpha(a)$) while the 
rest of the clusters of these two permutations are the same, 
$\Omega'_{\alpha} = \Omega_{\alpha} \; (\alpha \not= \alpha_{1}, \alpha_{2})$.
Correspondingly, this last contribution is proportional to 
$\delta(q_{\alpha_{1}} - q'_{\alpha_{2}}) \, \delta(q_{\alpha_{2}} - q'_{\alpha_{1}})\; (-1)^{n_{\alpha_{1}}}$.
In fact this situation with two equivalent contributions is the consequence of the symmetry
of the wave function $\Psi_{\bf q', n}^{(M)}({\bf x})$: 
the permutation of two momenta $q_{\alpha_{1}}$ and $q_{\alpha_{2}}$ belonging to 
the clusters which have the same numbers of particles, 
$n_{\alpha_{1}}=n_{\alpha_{2}}$ produces the factor $(-1)^{n_{\alpha_{1}}}$.
This is evident from the general expression for the wave function, eq.(\ref{A9}),
where the permutation of any two momenta $q_{\alpha_{1}}$ and $q_{\alpha_{2}}$ belonging to 
the clusters which have the same numbers of particles corresponds to the permutation of
$n$ columns of the matrix $\exp(i q_{a} x_{b})$. Therefore considering the clusters with equal numbers of particles as equivalent and restricting analysis to the sectors 
$q_{\alpha_{1}} < q_{\alpha_{2}}; \; q'_{\alpha_{1}} < q'_{\alpha_{2}}$
we find that the second contribution,  
$\delta(q_{\alpha_{1}} - q'_{\alpha_{2}}) \, \delta(q_{\alpha_{2}} - q'_{\alpha_{1}})$
is identically equal to zero, thus returning to the above result Eq.(\ref{B41}).

A generic eigenstate
$({\bf q}, {\bf n})$ with $M$ clusters could be specified in terms of the 
following set of parameters:
\begin{equation}
\label{B42}
({\bf q}, {\bf n}) \; = \;  \{
(\underbrace{q_{1}, m_{1}), ..., (q_{s_{1}}, m_{1})}_{s_{1}}; 
 \underbrace{(q_{s_{1}+1}, m_{2}), ..., (q_{s_{1}+s_{2}}, m_{2})}_{s_{2}}; 
\; .... \; ; 
 \underbrace{(q_{s_{1}+...+s_{k-1}+1}, m_{k}), ..., (q_{s_{1}+...+s_{k}}, m_{k})}_{s_{k}}\}
\end{equation}
where  $s_{1}+s_{2}+ ... +s_{k} = M$ and $k$ integers $\{ m_{i}\}$ ($ 1\leq k \leq M$)
are all supposed to be {\it different}: 
\begin{equation}
 \label{B43}
1 \leq m_{1} < m_{2} < ... < m_{k}
\end{equation}
Here the integer parameter $k$ denotes the number of different cluster types. For a given $k$
\begin{equation}
\label{B44}
\sum_{\alpha=1}^{M} n_{\alpha} \; = \; \sum_{i=1}^{k} s_{i} m_{i} \; =  \; N
\end{equation}
Due to the symmetry  with respect to the momenta permutations inside the subsets 
of equal $n$'s it is sufficient to consider the wave functions in the sectors
\begin{eqnarray}
\label{B45}
&&
q_{1} < q_{2} < ... < q_{s_{1}} \; ;\\
\nonumber
&&
q_{s_{1}+1} < q_{s_{1}+2} < ... < q_{s_{1}+s_{2}} \; ; \\
\nonumber
&&
................. \\
\nonumber
&&
q_{s_{1}+...+s_{k-1}+1} < q_{s_{1}+...+s_{k-1}+2} < ... < q_{s_{1}+...+s_{k-1}+s_{k}}
\end{eqnarray}
In this representation we again recover the above result Eq.(\ref{B41})

Finally, let us consider the overlap of two eigenstates described by two {\it different}
sets of integer parameters,  ${\bf n'} \not= {\bf n}$. In fact this situation is quite simple
because if the clusters of the two states are different from each other, it means
that in the summation over the pairs of permutations $P$ and $P'$ in Eq.(\ref{B31})
there exist no two permutations for which these two sets of clusters $\{\Omega_{\alpha}\}$ and 
$\{\Omega'_{\alpha}\}$ would coincide. Which, according to the above analysis, means that 
this expression is equal to zero. Note that the condition $M' \not= M$ automatically implies 
that ${\bf n'} \not= {\bf n}$. 

Thus we have proved that
\begin{eqnarray}
\nonumber
Q^{(M,M')}_{{\bf n},{\bf n'}}({\bf q}, {\bf q'}) &=& 
\big|C^{(M)}_{\bf q, n}\big|^{2}  \;
{\boldsymbol \delta}(M,M') \; \biggl(\prod_{\alpha=1}^{M} 
{\boldsymbol \delta}(n_{\alpha},n'_{\alpha}) \biggr)
\biggl(\prod_{\alpha=1}^{M} (2\pi) \delta(q_{\alpha}-q'_{\alpha}) \biggr)
\times
\\
\nonumber
\\
&& \times
 N! \; 
\prod_{\alpha=1}^{M} \Biggr[
\frac{n_{\alpha}\kappa}{(n_{\alpha}!)^{2} \kappa^{n_{\alpha}}} 
\Biggr]
\prod_{\alpha < \beta}^{M} \prod_{r=1}^{n_{\alpha}} \prod_{r'=1}^{n_{\beta}}
\Big| 
\bigl(q_{\alpha} - \frac{i\kappa}{2} n_{\alpha} \bigr) -
\bigl(q_{\beta}  - \frac{i\kappa}{2} n_{\beta}  \bigr)
+i\kappa \; (r - r' - 1)
\Big|^{2}  
\label{B46}
\end{eqnarray}
where the integer parameters $\{n_{\alpha}\}$ and $\{n'_{\alpha}\}$ are assumed to have the 
generic structure represented in Eqs.(\ref{B42})-(\ref{B44}), and the  momenta 
$\{q_{\alpha}\}$ and $\{q'_{\alpha}\}$ of the clusters with equal numbers of particles 
are restricted in the sectors, Eq.(\ref{B45}). 
According to Eq.(\ref{B46}), the orthonormality condition defines the normalization
constant 
\begin{equation}
   \label{B47}
  \big| C^{(M)}({\bf q,n})\big|^{2} = 
\frac{1}{N!} \; 
\biggl[
\prod_{\alpha=1}^{M} \frac{(n_{\alpha}!)^{2}\kappa^{n_{\alpha}} }
   {n_{\alpha} \kappa}
\biggr]
\prod_{\alpha < \beta}^{M} \prod_{r=1}^{n_{\alpha}} \prod_{r'=1}^{n_{\beta}}
\frac{1}{\Big|\bigl(q_{\alpha} - \frac{i\kappa}{2} n_{\alpha} \bigr) -
\bigl(q_{\beta}  - \frac{i\kappa}{2} n_{\beta}  \bigr)
+i\kappa \; (r - r' - 1)\Big|^{2}}
\end{equation}

\vspace{10mm}

{\center {\bf 4. Propagator}}

\vspace{3mm}

The time dependent solution $ \Psi({\bf x},t)$ of the imaginary-time 
Schr\"odinger equation
\begin{equation}
   \label{B48}
\beta \partial_t \Psi({\bf x}; t) \; = \;
\frac{1}{2}\sum_{a=1}^{N}\partial_{x_a}^2 \Psi({\bf x}; t)
  \; + \; \frac{1}{2}\kappa \sum_{a\not=b}^{N} \delta(x_a-x_b) \Psi({\bf x}; t)
\end{equation}
with the initial condition 
\begin{equation}
   \label{B49}
\Psi({\bf x}; 0) = \Pi_{a=1}^{N} \delta(x_a)
\end{equation}
 can be represented in terms of the 
linear combination of the eigenfunctions $\Psi_{\bf q, n}^{(M)}({\bf x})$, 
Eq.(\ref{B27}):
\begin{equation}
\label{B50}
\Psi({\bf x},t) \; = \; \sum_{M=1}^{N} \;
{\sum_{{\bf n}}}' \; 
\int ' {\cal D} {\bf q} \; \; 
 \Psi_{\bf q, n}^{(M)}({\bf x}) \Psi_{\bf q, n}^{(M)^{*}}({\bf 0}) \; 
\exp\bigl[-E_{M}({\bf q,n}) \; t \bigr]
\end{equation}
where the energy spectrum $E_{M}({\bf q,n})$ is given by Eq.(\ref{B28}).  
The summations over $n_{\alpha}$ are performed here in terms of the parameters $\{s_{i}, m_{i}\}$,
Eqs.(\ref{B42})-(\ref{B44}):
\begin{equation}
\label{B51}
{\sum_{{\bf n}}}' \; \equiv \;
\sum_{k=1}^{M} \; \; 
\sum_{s_{1}...s_{k}=1}^{\infty} \; \; 
\sum_{1 \leq m_{1} ... < m_{k}}^{\infty}
\; {\boldsymbol \delta}\biggl(\sum_{i=1}^{k} s_{i}, \; M\biggr)
\; \;{\boldsymbol \delta}\biggl(\sum_{i=1}^{k} s_{i} m_{i}, \; N\biggr)
\end{equation}
where ${\boldsymbol \delta}(n,l)$ is the Kronecker symbol,
and for simplicity (due to the presence of these Kronecker symbols)
the summations over $m_{i}$ and $s_{i}$
are extended to infinity. The symbol $\int '{\cal D} {\bf q}$ in Eq.(\ref{B50}) 
denotes the integration 
over $M$ momenta $q_{\alpha}$ in the sectors, Eq.(\ref{B45}). 

The replica partition function $Z(N,L)$ of the original directed polymer problem
is obtained via a particular choice of the final-point coordinates,
\begin{equation}
\label{B52}
Z(N,L) \; = \; \Psi({\bf 0};L) \; = \; 
\sum_{M=1}^{N} \;
{\sum_{{\bf n}}}' \; 
\int ' {\cal D} {\bf q} \; \; 
\bigl| \Psi_{\bf q, n}^{(M)}({\bf 0}) \bigr|^2 \; 
\exp\bigl[-E_{M}({\bf q,n}) \; L \bigr]
\end{equation}
According to Eq.(\ref{B27}), for $M \geq 2$,
\begin{eqnarray}
\nonumber
\Psi_{\bf q, n}^{(M)}({\bf 0}) 
&=& 
C^{(M)}_{\bf q, n}
{\sum_{P}}' (-1)^{[P]} 
\prod_{\substack{ a<b\\ \alpha(a)\not=\alpha(b) }}^{N}
\biggl[ 
  \Bigl(q_{\alpha(a)} - \frac{i\kappa}{2} \big[n_{\alpha(a)} + 1 - 2r(a)\big] \Bigr)
- \Bigl(q_{\alpha(b)} - \frac{i\kappa}{2} \big[n_{\alpha(b)} + 1 - 2r(b)\big] \Bigr)
\biggr] 
\\
\nonumber
\\
&=& 
C^{(M)}_{\bf q, n} \; \frac{N!}{n_{1}! n_{2}! ... n_{M}!} \;
\prod_{\alpha<\beta}^{M} \prod_{r=1}^{n_{\alpha}} \prod_{r'=1}^{n_{\beta}}
\biggl[ 
\Big(q_{\alpha} - \frac{i\kappa}{2} n_{\alpha} \Big) 
- \Big(q_{\beta} - \frac{i\kappa}{2} n_{\beta} \Big) 
+ i \kappa \bigl(r - r'\bigr)
\biggr]
\label{B53}
\end{eqnarray}
Substituting here the value of the normalization constant, Eq.(\ref{B47}), we get
\begin{equation}
 \label{B54}
\bigl| \Psi_{\bf q, n}^{(M)}({\bf 0}) \bigr|^2 \; = \; 
\frac{N! \kappa^{N}}{ \Bigl(\prod_{\alpha=1}^{M} \kappa n_{\alpha}\Bigr)} \; 
\prod_{\alpha<\beta}^{M} 
\frac{
\prod_{r=1}^{n_{\alpha}} \prod_{r'=1}^{n_{\beta}}
\Big|
\bigl(q_{\alpha} - \frac{i\kappa}{2} n_{\alpha} \bigr) -
\bigl(q_{\beta}  - \frac{i\kappa}{2} n_{\beta}  \bigr)
+i\kappa \; (r - r')
\Big|^{2}}{
\prod_{r=1}^{n_{\alpha}} \prod_{r'=1}^{n_{\beta}}
\Big|
\bigl(q_{\alpha} - \frac{i\kappa}{2} n_{\alpha} \bigr) -
\bigl(q_{\beta}  - \frac{i\kappa}{2} n_{\beta}  \bigr)
+i\kappa \; (r - r'-1)
\Big|^{2}}
\end{equation}
This expression can be essentially simplified. 
Shifting the product over $r'$ in the denominator by $1$ we obtain
\begin{equation}
 \label{B55}
\bigl| \Psi_{\bf q, n}^{(M)}({\bf 0}) \bigr|^2 \; = \; 
\frac{N! \kappa^{N}}{ \Bigl(\prod_{\alpha=1}^{M} \kappa n_{\alpha}\Bigr)} \; 
\prod_{\alpha<\beta}^{M} 
\frac{
\prod_{r=1}^{n_{\alpha}} 
\Big|
\bigl(q_{\alpha} - \frac{i\kappa}{2} n_{\alpha} \bigr) -
\bigl(q_{\beta}  - \frac{i\kappa}{2} n_{\beta}  \bigr)
+i\kappa \; (r - 1)
\Big|^{2}}{
\prod_{r=1}^{n_{\alpha}}
\Big|
\bigl(q_{\alpha} - \frac{i\kappa}{2} n_{\alpha} \bigr) -
\bigl(q_{\beta}  - \frac{i\kappa}{2} n_{\beta}  \bigr)
+i\kappa \; (r - n_{\beta} -1)
\Big|^{2}}
\end{equation}
Redefining the product parameter $r$ in the denominator, $ r \to n_{\alpha} + 1 - r$, 
and changing the obtained expression (under the modulus square) by its complex conjugate
we get
\begin{equation}
 \label{B56}
\bigl| \Psi_{\bf q, n}^{(M)}({\bf 0}) \bigr|^2 \; = \; 
\frac{N! \kappa^{N}}{ \Bigl(\prod_{\alpha=1}^{M} \kappa n_{\alpha}\Bigr)} \; 
\prod_{\alpha<\beta}^{M} 
\frac{
\prod_{r=1}^{n_{\alpha}} 
\Big|
\bigl(q_{\alpha} - \frac{i\kappa}{2} n_{\alpha} \bigr) -
\bigl(q_{\beta}  - \frac{i\kappa}{2} n_{\beta}  \bigr)
+i\kappa \; (r - 1)
\Big|^{2}}{
\prod_{r=1}^{n_{\alpha}}
\Big|
\bigl(q_{\alpha} - \frac{i\kappa}{2} n_{\alpha} \bigr) -
\bigl(q_{\beta}  - \frac{i\kappa}{2} n_{\beta}  \bigr)
+i\kappa \; r
\Big|^{2}}
\end{equation}
Shifting now the product over $r$ in the numerator by $1$ we finally obtain
\begin{equation}
 \label{B57}
\bigl| \Psi_{\bf q, n}^{(M)}({\bf 0}) \bigr|^2 \; = \; 
\frac{N! \kappa^{N}}{ \Bigl(\prod_{\alpha=1}^{M} \kappa n_{\alpha}\Bigr)} \; 
\prod_{\alpha<\beta}^{M} 
\frac{\big|q_{\alpha}-q_{\beta} -\frac{i\kappa}{2}(n_{\alpha}-n_{\beta})\big|^{2}}{
      \big|q_{\alpha}-q_{\beta} -\frac{i\kappa}{2}(n_{\alpha}+n_{\beta})\big|^{2}}
\end{equation}
For $M=1$, according to Eqs.(\ref{B1}) and (\ref{B11}),
\begin{equation}
\label{B58}
\big|\Psi_{q}^{(1)}({\bf 0})\big|^{2} \; = \; \frac{\kappa^N N!}{\kappa N}
\end{equation}
Since the function $f\bigl({\bf q}, {\bf n}\bigr) = \bigl| \Psi_{\bf q, n}^{(M)}({\bf 0}) \bigr|^2 \; 
\exp\bigl[-E_{M}({\bf q,n}) \; L \bigr]$ in Eq.(\ref{B52}) is symmetric 
with respect to permutations of all its $M$ pairs of arguments $(q_{\alpha}, n_{\alpha})$
the integrations over $M$ momenta $q_{\alpha}$ can be extended beyond the sector defined in Eq.(\ref{B45})
for the whole space $R_{M}$. As a consequence, there is no need 
to distinguish equal and different $n_{\alpha}$'s any more, and instead of Eq.(\ref{B51}), we can sum
over $M$ integer parameters $n_{\alpha}$ with the only constrain, Eq.(\ref{B17})
(note that this kind of simplifications holds only for the specific "zero final-point" object
$\Psi({\bf 0};t)$, Eq.(\ref{B52}), and {\it not} for the general propagator
$\Psi({\bf x};t)$, Eq.(\ref{B50}) containing $N$ arbitrary coordinates $x_{1}, ..., x_{N}$).
Thus, instead of Eq.(\ref{B52}) we get
\begin{equation}
\label{B59}
Z(N,L) \; = \;  
\sum_{M=1}^{N} \; 
\frac{1}{M!}
\biggl[
\prod_{\alpha=1}^{M} 
\int_{-\infty}^{+\infty} \; \frac{dq_{\alpha}}{2\pi} 
\sum_{n_{\alpha}=1}^{\infty}
\biggr] \;
{\boldsymbol \delta}\biggl(\sum_{\alpha=1}^{M} n_{\alpha}, \; N \biggr) \;
\big| \Psi_{\bf q, n}^{(M)}({\bf 0}) \big|^2 \; 
\mbox{\LARGE e}^{-E_{M}({\bf q,n}) L}
\end{equation}
Substituting here Eqs.(\ref{B28}), (\ref{B57}) and (\ref{B58}) we get the following sufficiently
compact representation for the replica partition function:
\begin{eqnarray}
\nonumber
Z(N.L) &=& 
N! \; \kappa^{N} \biggl\{
 \int_{-\infty}^{+\infty} \frac{dq}{2\pi\kappa N} \;
\exp\Bigl[
-\frac{L}{2\beta} N q^{2} + \frac{\kappa^{2}L}{24\beta} (N^{3} -N) 
\Bigr] \; +
\\
\nonumber
\\
\nonumber
&+& 
\sum_{M=2}^{N} \frac{1}{M!} 
\biggl[
\prod_{\alpha=1}^{M}
\sum_{n_{\alpha}=1}^{\infty}
\int_{-\infty}^{+\infty} \frac{d q_{\alpha}}{2\pi\kappa n_{\alpha}} 
\biggr]
\;{\boldsymbol \delta}\biggl(\sum_{\alpha=1}^{M} n_{\alpha}, \; N \biggr) 
\prod_{\alpha<\beta}^{M} 
\frac{\big|q_{\alpha}-q_{\beta} -\frac{i\kappa}{2}(n_{\alpha}-n_{\beta})\big|^{2}}{
      \big|q_{\alpha}-q_{\beta} -\frac{i\kappa}{2}(n_{\alpha}+n_{\beta})\big|^{2}} 
\times
\\
&\times&
\exp\Bigl[
-\frac{L}{2\beta}\sum_{\alpha=1}^{M} n_{\alpha} q_{\alpha}^{2} + 
\frac{\kappa^{2}L}{24\beta} \sum_{\alpha=1}^{M} (n_{\alpha}^{3} - n_{\alpha})
\Bigr] 
\biggr\}
\label{B60}
\end{eqnarray}
The first term in the above expression is the contribution of the ground state $(M=1)$,
while  the next terms $(M \geq 2)$ are the contributions of the rest of the energy
spectrum. After simple algebra the above replica partition function can be represented as follows:
\begin{equation}
   \label{B61}
 Z(N,L) \; =  \mbox{\LARGE e}^{-\beta N L f_{0}} \;\;
\tilde{Z}(N,\lambda) 
\end{equation}
where $f_{0} = \frac{1}{24}\beta^4 u^2 - \frac{1}{\beta L} \ln(\beta^{3} u)$, 
and
\begin{eqnarray}
\nonumber
\tilde{Z}(N.L) &=& 
N! \; \biggl\{
 \int_{-\infty}^{+\infty} \frac{dq}{2\pi\kappa N} \;
\exp\Bigl[
-\frac{L}{2\beta} N q^{2} + \frac{\kappa^{2}L}{24\beta} N^{3} 
\Bigr]
\; +
\\
\nonumber
\\
\nonumber
&+& 
\sum_{M=2}^{N} \frac{1}{M!} 
\biggl[
\prod_{\alpha=1}^{M}
\sum_{n_{\alpha}=1}^{\infty}
\int_{-\infty}^{+\infty} \frac{d q_{\alpha}}{2\pi\kappa n_{\alpha}} 
\biggr]
\;{\boldsymbol \delta}\biggl(\sum_{\alpha=1}^{M} n_{\alpha}, \; N \biggr) 
\prod_{\alpha<\beta}^{M} 
\frac{\big|q_{\alpha}-q_{\beta} -\frac{i\kappa}{2}(n_{\alpha}-n_{\beta})\big|^{2}}{
      \big|q_{\alpha}-q_{\beta} -\frac{i\kappa}{2}(n_{\alpha}+n_{\beta})\big|^{2}} 
\times
\\
&\times&
\exp\Bigl[
-\frac{L}{2\beta}\sum_{\alpha=1}^{M} n_{\alpha} q_{\alpha}^{2} + 
\frac{\kappa^{2}L}{24\beta} \sum_{\alpha=1}^{M} n_{\alpha}^{3}
\Bigr] 
\biggr\}
\label{B62}
\end{eqnarray}

\vspace{10mm}

\begin{center}

\appendix{\Large Appendix C}

\vspace{5mm}

 {\bf \large The Airy function integral relations}

\end{center}

\newcounter{C}
\setcounter{equation}{0}
\renewcommand{\theequation}{C.\arabic{equation}}

\vspace{5mm}

The Airy function $\Ai(x)$ is the solution of the differential equation
\begin{equation}
\label{C1}
y''(x) \; = \; x\, y(x)
\end{equation}
with the boundary condition $y(x\to +\infty) = 0$. At $x\to +\infty$ this function
goes to zero exponentially fast
\begin{equation}
\label{C2}
\Ai(x\to +\infty) \; \simeq \; \frac{1}{2\sqrt{\pi} x^{1/4}} \exp\Bigl(-\frac{2}{3} x^{3/2}\Bigr)
\end{equation}
while at $x\to -\infty$ it oscillates and decays much more slowly:
\begin{equation}
\label{C3}
\Ai(x\to -\infty) \; \simeq \; \frac{1}{\sqrt{\pi} |x|^{1/4}} 
\sin\Bigl(\frac{2}{3} |x|^{3/2} + \frac{1}{4}\pi\Bigr)
\end{equation}
The Airy function can also be represented in the integral form:
\begin{equation}
\label{C4}
\Ai(x) \; = \; \int_{{\cal C}} \frac{dz}{2\pi i} \; \exp\Bigl(\frac{1}{3} z^{3} - z x\Bigr)
\end{equation}
where the integration path in the complex plane starts at a point at infinity with the argument
$-\pi/2 < \theta_{(-)} < -\pi/3$ and ends up at a point at infinity with the argument
$\pi/3 < \theta_{(+)} < \pi/2$. Choosing the argument of the staring point 
$\theta_{(-)} = -\pi/2 + \epsilon$ and that of the ending point 
$\theta_{(+)} = \pi/2 - \epsilon$ where the positive parameter $\epsilon \to 0$ is introduced 
just to provide the convergence of the integration, the integration 
path in eq.(\ref{C4}) can be chosen to be coinciding with the imaginary axes $z = i y$.

Just like in the well known Hubbard-Stratonovich transformation the Gaussian function
is used to linearize quadratic expressions in the exponential,
\begin{equation}
\label{C5}
\exp\Bigl(\frac{1}{2} F^{2} \Bigr) \; = \; 
\int_{-\infty}^{+\infty} \frac{dx}{\sqrt{2\pi}} \; 
\exp\Bigl(-\frac{1}{2} x^{2} \Bigr) \,
\exp\bigl( F x\bigr)
\end{equation}
the Airy function can be used to linearize the {\it cubic} exponential terms:
\begin{equation}
\label{C6}
\exp\Bigl(\frac{1}{3} F^{3} \Bigr) \; = \; 
\int_{-\infty}^{+\infty} \, dx \; \Ai(x) \,
\exp\bigl( F x \bigr)
\end{equation}
where the quantity $F$ in eq.(\ref{C6}) is assumed to be non negative. One can easily prove this 
relation using integral representation, eq.(\ref{C4}), in which the integration path coincides
with the imaginary axes, $z = i y$, and the quantity $F$ is first taken to be pure imaginary,
$F \to i F$. In this case the integration over $x$ results in the factor $\delta\bigl(F-y\bigr)$.
Further trivial integration over $y$ yields the result 
$\exp\bigl(- i F^{3}/3 \bigr)$. Performing the  analytic continuation of this expression
$F \to - i F$ one get the relation eq.(\ref{C6}).  

In this appendix two other integral relations with the Airy function will be proved.
Namely:
\begin{equation}
\label{C7}
I_{1} \; \equiv \; \int_{-\infty}^{+\infty} dp \;
\Ai\bigl(p^{2} + \omega_{1} + \omega_{2}\bigr) \; 
\exp\bigl[i p ( \omega_{1} - \omega_{2}) \bigr]
\; = \; 
2^{2/3} \pi \Ai\bigl(2^{1/3} \omega_{1}\bigr)  \Ai\bigl(2^{1/3} \omega_{2}\bigr) 
\end{equation}
and 
\begin{equation}
\label{C8}
I_{2} \; \equiv \; \int_{0}^{\infty} dy \; 
\Ai\bigl(y + \omega_{1}\bigr) \Ai\bigl(y + \omega_{2}\bigr) 
\; = \; 
\frac{\Ai\bigl(\omega_{1}\bigr) \Ai'\bigl(\omega_{2}\bigr) \; - \; 
\Ai'\bigl(\omega_{1}\bigr) \Ai\bigl(\omega_{2}\bigr)}{\omega_{1} - \omega_{2}}
\end{equation}

Using the integral representation of the Airy function, eq.(\ref{C4}), we get
\begin{equation}
\label{C9}
I_{1} \; = \; \int_{-\infty}^{+\infty} dp 
\int_{{\cal C}}\frac{dz}{2\pi i}\;
\exp\Bigl(\frac{1}{3} z^{3} - p^{2} z - \omega_{1} z - \omega_{2} z 
+ i p \omega_{1} - i p \omega_{2} \Bigr)
\end{equation}
Denoting $z \equiv z_{1}$ and $ip \equiv z_{2}$ the above integrals can be 
represented as follows
\begin{equation}
\label{C10}
I_{1} \; = \; 2\pi \int\int_{{\cal C}} \frac{dz_{1} dz_{2}}{(2\pi i)^{2}}\;
\exp\Bigl(\frac{1}{3} z_{1}^{3} + z_{1} z_{2}^{2} 
- \omega_{1} (z_{1} - z_{2}) - \omega_{2} (z_{1} + z_{2}) \Bigr)
\end{equation}
where the integration path ${\cal C}$ coincides with the imaginary axes.
Introducing new integration variables,
\begin{eqnarray}
\nonumber
z_{1} - z_{2} &=& \xi
\\
z_{1} + z_{2} &=& \eta
\label{C11}
\end{eqnarray}
we obtain
\begin{equation}
\label{C12}
I_{1} \; = \; \pi 
\int_{{\cal C}} \frac{d\xi}{2\pi i}\;
\int_{{\cal C}} \frac{d\eta}{2\pi i}
\exp\Bigl(\frac{1}{6} \xi^{3} + \frac{1}{6} \eta^{3} 
- \omega_{1} \xi - \omega_{2} \eta \Bigr)
\end{equation}
Redefining, $\xi \to 2^{1/3} \xi$ and $\eta \to 2^{1/3} \eta$ we finally get
\begin{equation}
\label{C13}
I_{1} \; = \; 2^{2/3} \pi 
\int_{{\cal C}} \frac{d\xi}{2\pi i}\;
\exp\Bigl(\frac{1}{3} \xi^{3} - 2^{1/3} \omega_{1} \xi \Bigr)
\int_{{\cal C}} \frac{d\eta}{2\pi i}
\exp\Bigl(\frac{1}{3} \eta^{3} - 2^{1/3} \omega_{2} \eta \Bigr) 
\; = \; 
2^{2/3} \pi \Ai\bigl(2^{1/3} \omega_{1}\bigr)  \Ai\bigl(2^{1/3} \omega_{2}\bigr) 
\end{equation}
which proves the relation, eq.(\ref{C7}).

To prove the relation, eq.(\ref{C8}), it is sufficient to take into account that
the Airy function satisfies the differential equation (\ref{C1}). 
Substituting
\begin{equation}
\label{C14}
\Ai\bigl(y + \omega\bigr) \; = \; \frac{1}{y+\omega} \, \Ai''\bigl(y + \omega\bigr)
\end{equation}
into eq.(\ref{C8}) we find
\begin{eqnarray}
\nonumber
I_{2} &=&
\int_{0}^{\infty} dy \, \frac{1}{(y + \omega_{1})(y + \omega_{2})} \, 
\Ai''\bigl(y + \omega_{1}\bigr) \, \Ai''\bigl(y + \omega_{2} \bigr)
\\
\nonumber
\\
\nonumber
&=&
\frac{1}{\omega_{1}-\omega_{2}} \,
\int_{0}^{\infty} dy \,
\biggl[\frac{1}{y+\omega_{2}} - \frac{1}{y+\omega_{1}} \biggr] \, 
\Ai''\bigl(y + \omega_{1}\bigr) \, \Ai''\bigl(y + \omega_{2} \bigr)
\\
\nonumber
\\
&=& 
\frac{1}{\omega_{1}-\omega_{2}} \,
\int_{0}^{\infty} dy \,
\biggl[
\frac{\Ai''(y+\omega_{2})}{y+\omega_{2}} \, \Ai''(y+\omega_{1}) \; - \; 
\frac{\Ai''(y+\omega_{1})}{y+\omega_{1}} \, \Ai''(y+\omega_{2})
\biggr]
\label{C15}
\end{eqnarray}
Substituting here
 $\Ai''(y+\omega_{2}) = (y+\omega_{2}) \Ai(y+\omega_{2})$
for the first term and 
 $\Ai''(y+\omega_{1}) = (y+\omega_{1}) \Ai(y+\omega_{1})$
for the second term we get
\begin{equation}
\label{C16}
I_{2} \; = \; 
\frac{1}{\omega_{1}-\omega_{2}} \,
\biggl[
\int_{0}^{\infty} dy \,
\Ai''(y+\omega_{1}) \Ai(y+\omega_{2}) \; - \; 
\int_{0}^{\infty} dy \,
\Ai''(y+\omega_{2}) \Ai(y+\omega_{1})
\biggr]
\end{equation}
Simple integration by parts eventually yields
\begin{equation}
\label{C17}
I_{2} \; = \; 
\frac{1}{\omega_{1}-\omega_{2}} \,
\Bigl[ - \Ai'(\omega_{1}) \Ai(\omega_{2}) + \Ai'(\omega_{2}) \Ai(\omega_{1}) 
\Bigr]
\end{equation}
which proves the relation, eq.(\ref{C8}).

\vspace{10mm}

\begin{center}

\appendix{\Large Appendix D}

\vspace{5mm}

 {\bf \large Fredholm determinant with the Airy kernel and the Tracy-Widom distribution}

\end{center}

\newcounter{D}
\setcounter{equation}{0}
\renewcommand{\theequation}{D.\arabic{equation}}

\vspace{5mm}

In simplified terms the Fredholm determinant $\det \bigl(1 - \hat{K}\bigr)$
can be defined as follows 
(for the strict mathematical definition see e.g. \cite{Mehta}):
\begin{equation}
\label{D1}
\det \bigl(1 - \hat{K}\bigr) \; = \; 
1 \; + \; \sum_{n=1}^{\infty} \frac{(-1)^{n}}{n!}
\int\int ...\int_{a}^{b} dt_{1} dt_{2} ... dt_{n} \; 
\det\Bigl[K(t_{i}, t_{j})\Bigr]_{(i,j) = 1, ..., n}
\end{equation}
where the {\it kernel} $\hat{K} \equiv K(t,t')$ is a function of two
variables defined in a region $a \leq (t,t') \leq b$. Equivalently 
the Fredholm determinant can also be represented in the 
exponential form
\begin{equation}
\label{D2}
\det \bigl(1 - \hat{K}\bigr) \; = \; 
\exp\biggl[-\sum_{n=1}^{\infty} \frac{1}{n} \, \mbox{Tr} \, \hat{K}^{n} \biggr]
\end{equation}
where
\begin{equation}
\label{D3}
\mbox{Tr} \, \hat{K}^{n} \; \equiv \;
\int\int ...\int_{a}^{b} dt_{1} dt_{2} ... dt_{n} \;
K(t_{1}, t_{2}) K(t_{2}, t_{3}) ... K(t_{n}, t_{1})
\end{equation}

In this Appendix the original derivation of  Tracy and Widom \cite{Tracy-Widom} will be repeated 
in simple terms to demonstrate that the function $F_{2}(s)$
defined as the Fredholm determinant with the Airy kernel can be expressed in terms of the 
solution of the Panlev\'e II differential equation, namely
\begin{equation}
 \label{D4}
 F_{2}(s) \equiv \det\bigl[1 - \hat{K}_{A}\bigr] \; = \; 
\exp\biggl[
-\int_{s}^{\infty} dt \, (t-s) q^{2}(t) 
\biggr]
\end{equation}
where $\hat{K}_{A}$ is the Airy kernel defined on semi-infinite interval $[s, \infty)$:  
\begin{equation}
 \label{D5}
K_{A}(t_{1},t_{2}) \; = \; 
\frac{\Ai(t_{1}) \Ai'(t_{2}) - \Ai'(t_{1}) \Ai(t_{2})}{t_{1} - t_{2}}
\end{equation}
and the function $q(t)$ is the solution of the Panlev\'e II differential equation,
\begin{equation}
 \label{D6}
q'' = t q + 2 q^{3}
\end{equation}
with the boundary condition, $q(t\to +\infty) \sim \Ai(t)$.

\vspace{5mm}

Let us introduce a new function $R(t)$ such that
\begin{equation}
 \label{D7}
 F_{2}(s) \; = \; 
\exp\biggl[
-\int_{s}^{\infty} dt R(t) 
\biggr]
\end{equation}
or, according to the definition, Eq.(\ref{D4}),
\begin{equation}
 \label{D8}
R(s) \; = \; \frac{d}{ds} \ln\Bigl[\det\bigl(1 - \hat{K}_{A}\bigr) \Bigr]
\end{equation}
Here the logarithm of the determinant can be expressed in terms of the trace:
\begin{eqnarray}
 \label{D9}
\ln\Bigl[\det\bigl(1 - \hat{K}_{A}\bigr) \Bigr] 
&=&
- \sum_{n=1}^{\infty} \frac{1}{n} \, Tr \, \hat{K}_{A}^{n} 
\\
\nonumber
\\
\nonumber
&\equiv& 
- \sum_{n=1}^{\infty} \frac{1}{n}
\int_{s}^{\infty} dt_{1} \int_{s}^{\infty} dt_{2} \, ... \, \int_{s}^{\infty} dt_{n} \; 
 K_{A}(t_{1},t_{2}) K_{A}(t_{2},t_{3}) \, ... \, K_{A}(t_{n},t_{1}) 
\end{eqnarray}
Taking derivative of this expression we gets
\begin{eqnarray}
 \label{D10}
R(s) &=& - \int_{s}^{\infty} dt \, \bigl(1 - \hat{K}_{A}\bigr)^{-1} (s,t) \, K_{A}(t,s)
\\
\nonumber
\\
\nonumber
&\equiv& 
 - K_{A}(s,s) 
- \sum_{n=2}^{\infty} 
\int_{s}^{\infty} dt_{1} \int_{s}^{\infty} dt_{2} \, ... \, \int_{s}^{\infty} dt_{n-1} \; 
K_{A}(s,t_{1}) K_{A}(t_{1},t_{2}) \, ... \, K_{A}(t_{n-1},s) 
\end{eqnarray}
Substituting here the integral representation of the Airy kernel, Eq.(\ref{D5}),
\begin{equation}
 \label{D11}
K_{A}(t_{1},t_{2}) \; = \; 
\int_{0}^{\infty} dz \Ai(t_{1} + z) \, \Ai(t_{2} + z)
\end{equation}
after some efforts in simple algebra one gets
\begin{equation}
 \label{D12}
R(s) \; = \; 
\int_{s}^{\infty} dt_{1} \int_{s}^{\infty} dt_{2} \, 
\Ai(t_{1}) \, \bigl(1 - \hat{K}_{A}\bigr)^{-1} (t_{1},t_{2}) \, \Ai(t_{2})
\end{equation}
Taking the derivative of this expression and applying some more efforts in 
 slightly more complicated algebra, we obtain
\begin{equation}
 \label{D13}
\frac{d}{ds} R(s) \; = \; - q^{2}(s)
\end{equation}
where
\begin{equation}
 \label{D14}
q(s) \; = \; \int_{s}^{\infty} dt \, \bigl(1 - \hat{K}_{A}\bigr)^{-1} (s,t) \, \Ai(t)
\end{equation}
According to Eq.(\ref{D13}),
\begin{equation}
 \label{D15}
 R(s) \; = \; \int_{s}^{\infty} dt \, q^{2}(t)
\end{equation}
Let us introduce two more functions
\begin{eqnarray}
 \label{D16}
v(s) &=& \int_{s}^{\infty} dt_{1} \int_{s}^{\infty} dt_{2} \, 
\Ai(t_{1}) \, \bigl(1 - \hat{K}_{A}\bigr)^{-1} (t_{1},t_{2}) \, \Ai'(t_{2})
\\
\nonumber
\\
\label{D17}
p(s) &=& \int_{s}^{\infty} dt \, \bigl(1 - \hat{K}_{A}\bigr)^{-1} (s,t) \, \Ai'(t)
\end{eqnarray}
Taking derivatives of the above three functions $q(s)$, $v(s)$ and $p(s)$, 
Eqs.(\ref{D14}), (\ref{D16}) and (\ref{D17}),
after somewhat painful algebra one finds the following three relations:
 \begin{eqnarray}
 \label{D18}
q' &=& p \; - \; R \, q
\\
\label{D19}
p' &=& s \, q \; - \; p \, R \; - \; 2 q \, v
\\
\label{D20}
v' &=& - p \, q
\end{eqnarray}
Taking derivative of the combination $\bigl(R^{2} - 2v\bigr)$ and using Eqs.(\ref{D13}) and (\ref{D20}),
we get
\begin{equation}
 \label{D21}
\frac{d}{ds} \bigl(R^{2} - 2v\bigr) \; = \; 2q \, (p \; - \; R\, q)
\end{equation}
On the other hand, multiplying Eq.(\ref{D18}) by $2q$ we find
\begin{equation}
 \label{D22}
\frac{d}{ds} q^{2} \; = \; 2q \, (p \; - \; R\, q)
\end{equation}
Comparing Eqs.(\ref{D21}) and (\ref{D22}) and taking into account that 
the value of all the above functions at $s\to\infty$ is zero, we obtain the following relation
\begin{equation}
 \label{D23}
R^{2} - 2v \; = \; q^{2}
\end{equation}
Finally, taking the derivative of Eq.(\ref{D18}) and using 
Eqs.(\ref{D13}), (\ref{D18}), (\ref{D19})
and (\ref{D23}) we easily find
\begin{equation}
 \label{D24}
q'' \; = \; 2q^{3} \; + \; s q
\end{equation}
which is the special case of the Panlev\'e II differential equation 
\cite{Panleve,Clarkson,Iwasaki}.
Thus, substituting Eq.(\ref{D15}) into Eq.(\ref{D7}) we obtain Eq.(\ref{D4}).

In the limit $s\to\infty$ the function $q(s)$, according to its definition, eq.(\ref{D14}),
must go to zero, and in this case Eq.(\ref{D24})
turns into the Airy function equation, $q'' = s q$. Thus
\begin{equation}
 \label{D27}
q(s\to\infty) \; \simeq \; \Ai(s) \; \sim \; \exp\Bigl[-\frac{2}{3} s^{3/2}\Bigr]
\end{equation}
It can be proved \cite{Hastings} that in the opposite limit, $s\to -\infty$, the asymptotic form 
of the solution of the Panleve\'e equation (\ref{D24}) (which has the right tail Airy function
limit, Eq.(\ref{D27})) is
\begin{equation}
 \label{D28}
q(s\to -\infty) \; \simeq \; \sqrt{-\frac{1}{2}s}
\end{equation}

\vspace{5mm}

The Tracy-Widom distribution function $P_{TW}(t)$ is defined by the 
function $F_{2}(s)$ as follows.
The function $F_{2}(s)$ gives the probability 
that a random quantity $t$ described by a probability distribution functions
$P_{TW}(t)$ has the value less than a given parameter $s$:
\begin{equation}
 \label{D25}
F_{2}(s) \; = \; \int_{-\infty}^{s} dt \, P_{TW}(t)
\end{equation}
Taking the derivative of this relation and substituting here the result, Eq.(\ref{D4}), we find
\begin{equation}
 \label{D26}
P_{TW}(s) \; = \; 
\exp\biggl[
-\int_{s}^{\infty} dt \, (t-s) q^{2}(t) 
\biggr] \times \int_{s}^{\infty} dt \, q^{2}(t)
\end{equation}
where the function $q(s)$ is the solution of the differential equation (\ref{D24})

Substituting the above two asymptotics into Eq.(\ref{D26}), 
we can estimate the asymptotic behavior
for the right and the left tails of the TW probability distribution function:
\begin{eqnarray}
 \label{D29}
P_{TW}(s\to +\infty) &\sim&
 \exp\Bigl[-\frac{4}{3} s^{3/2}\Bigr]
\\
\nonumber
\\
\label{D30}
P_{TW}(s\to -\infty) &\sim&
 \exp\Bigl[-\frac{1}{12} |s|^{3}\Bigr]
\end{eqnarray}

%\newpage

%%%%%%%%%%%%%%%%%%%%%%%%%%%%%%%%%%%%%%%%%%%%%%%%%%%%%%%%%%%%%%%%%%%%%%%%%%%%%%%%%%%
%%%%%%%%%%%%%%%%%%%%%%%%%%%%%%%%%%%%%%%%%%%%%%%%%%%%%%%%%%%%%%%%%%%%%%%%%%%%%%%%%%%


\begin{thebibliography}{99}


%%%% Intro

\bibitem{Tracy-Widom} C.A.\ Tracy and H.\ Widom,
   Commun.\ Math.\ Phys.\ {\bf 159}, 151 (1994)




\bibitem{LIS} J.\ Baik, P.A.\ Deift and K.\ Johansson, 
   J.\ Amer.\ Math.\ Soc.\ {\bf 12}, 1119 (1999)

\bibitem{DP_johansson} K.\ Johansson, 
   Comm.\ Math.\ Phys.\ {\bf 209}, 437 (2000)

\bibitem{PNG_Spohn} M.\ Pr\"ahofer and H.\ Spohn,
   Phys.\ Rev.\ Lett.\ {\bf 84}, 4882 (2000).

\bibitem{oriented_boiling} J.\ Gravner, C.A.\ Tracy and H.\ Widom, 
   J.\ Stat.\ Phys.\ {\bf 102}, 1085 (2001)

\bibitem{ballistic_decomposition} S.N.\ Majumdar and S.\ Nechaev, 
   Phys.\ Rev.\ E {\bf 69}, 011103 (2004)

\bibitem{LCS} S.N.\ Majumdar and S.\ Nechaev, 
   Phys.\ Rev.\ E {\bf 72}, 020901(R) (2005)

\bibitem{KPZ} M.Kardar, G.Parisi,Y-C.Zhang, 
   Phys.\ Rev.\ Lett.\ {\bf 56}, 889 (1986)

\bibitem{KPZ-TW1} T.Sasamoto and H.Spohn, 
   J. Stat. Phys. {\bf 140}, 209 (2010); arXiv:1002.1873; 
   Nucl. Phys. B {\bf 834}, 523 (2010); arXiv:1002.1879; 
   Phys. Rev. Lett. {\bf 104}, 230602 (2010); arXiv:1002.1883

\bibitem{KPZ-TW2}  G.Amir, I.Corwin and J.Quastel, 
   arXiv:1003.0443

\bibitem{Dotsenko1} V.Dotsenko and B.Klumov, 
   J.Stat.Mech. P03022 (2010)

\bibitem{Dotsenko2} V.Dotsenko,
    EPL, {\bf 90},20003 (2010)

\bibitem{Dotsenko3} V.Dotsenko, 
   J.Stat.Mech. P07010 (2010)  

\bibitem{LeDoussal} P.Calabrese, P. Le Doussal and A.Rosso
      EPL, {\bf 90},20002 (2010);
    

 


\bibitem{hh_zhang_95} T.\ Halpin-Healy and Y-C.\ Zhang, 
   Phys.\ Rep.\ {\bf 254}, 215 (1995).

\bibitem{lemerle_98} S.\ Lemerle, J.\ Ferr\'e, C.\ Chappert,
   V.\ Mathet, T.\ Giamarchi, and P.\ Le Doussal,
   Phys.\ Rev.\ Lett.\ {\bf 80}, 849 (1998).

\bibitem{blatter_94} G.\ Blatter, M.V.\ Feigel'man, V.B.\ 
   Geshkenbein, A.I.\ Larkin, and V.M.\ Vinokur, 
   Rev.\ Mod.\ Phys.\  {\bf 66}, 1125 (1994).

\bibitem{wilkinson_83} D.\ Wilkinson and J.F.\ Willemsen,
   J.\ Phys.\ A {\bf 16}, 3365 (1983).

\bibitem{burgers_74} J.M.\ Burgers, {\it The Nonlinear 
   Diffusion Equation} (Reidel, Dordrecht, 1974).


%%%% Intro I.1

\bibitem{Ulam} S.M.Ulam, 
      {\it Modern Mathematics for the Engineers}, ed. by E.F.Beckenbach
      (McGraw-Hill, New York, 1961)

\bibitem{Vershik-Kerov} A.M.Vershik and S.V.Kerov,
     Sov.Math.Dokl., {\bf 18}, 527 (1977)

\bibitem{Aldous} D.Aldous, P.Diaconis, \
     Bull.\ Amer.\ Math.\ Soc.\ {\bf 36}, 413 (1999)

\bibitem{Ferrari} P.L.Ferrari
     {\it Shape Fluctuations of crystal Facets and Surface Growth in 
      One Dimension}, PhD Thesis, Technische Universit\"at M\"unchen (2004)

\bibitem{takeuchi} Kazumasa A. Takeuchi and Masaki Sano,
        Phys.\ Rev.\ Lett.\ {\bf 104}, 230601 (2010)


\bibitem{Krug} J.Krug, P.Meakin and T.Halpin-Healy,
     Phys.Rev. A {\bf 45}, 638 (1992)



\bibitem{hhf_85} D.A.\ Huse, C.L.\ Henley, and D.S.\ Fisher,
    Phys.\ Rev.\ Lett.\ {\bf 55}, 2924 (1985).

\bibitem{numer1} D.A.\ Huse and C.L.\ Henley,
    Phys.\ Rev.\ Lett. {\bf 54}, 2708 (1985).

\bibitem{numer2} M.\ Kardar and Y-C.\ Zhang,
    Phys.\ Rev.\ Lett.\  {\bf 58}, 2087 (1987).


\bibitem{kardar_87} M.\ Kardar,
   Nucl.\ Phys.\ {\bf B 290}, 582 (1987).


%%%
\bibitem{ufn} V.S.Dotsenko, 
              Physics Uspekhi, {\bf 38}, 457 (1995)

\bibitem{book} Victor Dotsenko, 
               {\it Introduction to the Replica Theory of Disordered statistical Systems},
               Cambridge University Press, (2001)


\bibitem{zirnbauer1} J.J.M.Verbaarschot and M.R.Zirnbauer,
         J.Phys A: Math. Gen. {\bf 18}, 1093 (1985)


\bibitem{replicas} Victor Dotsenko,
          {\it One more discussion of the replica trick}, ArXiv:1010.3913 (2010)


\bibitem{REM} B. Derrida, 
         Phys. Rev. B {\bf 24}, 2613 (1981). 

%%%%%

\bibitem{Wigner} E.P.\ Wigner, 
   Proc.\ Cambridge Philos.\ Soc., {\bf 47}, 790 (1951)

\bibitem{Tracy-Widom2} C.A.\ Tracy and H.\ Widom,
   Commun.\ Math.\ Phys.\ {\bf 177}, 727 (1996)




\bibitem{Panleve} P.Panleve\'e, 
   {\it Sur les \'equation diff\'erentielles du second odre et odre sup\'erieur dont 
        int\'egrale g\'en\'erale est uniforme}. 
   Acta.\, Math. {\bf 25}, 1 (1902)

\bibitem{Clarkson} P.A.Clarkson,
   J.\ Comp.\ Appl.\ Math.\ {\bf 153}, 127 (2003)






\bibitem{Lieb-Liniger} E.H.\ Lieb and W.\ Liniger, 
     Phys.\ Rev.\  {\bf 130}, 1605 (1963)

\bibitem{bogolubov} V.E.\ Korepin, N.M.\ Bogoliubov, and A.G.\ Izergin, 
     {\it Quantum inverse scattering method and correlation functions}\\
     (Cambridge Univ. Press, Cambridge, 1993)

\bibitem{gaudin} M.\ Gaudin, {\it La fonction d'onde de Bethe}, 
     (Paris, Masson, 1983)

\bibitem{McGuire} J.B.\ McGuire, 
   J.\ Math.\ Phys.\ {\bf 5}, 622 (1964).

\bibitem{Yang} C.N.\ Yang,
     Phys.\ Rev.\  {\bf 168}, 1920 (1968)

\bibitem{Takahashi} M.\ Takahashi, {\it Thermodynamics of 
   one-dimensional solvable models} (Cambridge University Press, 1999).

\bibitem{Calabrese} P.\ Calabrese and J.-S.\ Caux,
   Phys.\ Rev.\ Lett.\ {\bf 98}, 150403 (2007).

\bibitem{Mehta} M.L.Mehta, {\it Random Matrices} (Elsevier, Amsterdam 2004)



%%%%%%%%%%%%%%%%%%%%%%%%%%%%%%%%%%%


\bibitem{Medina_93} E.\ Medina and M.\ Kardar,
   J.\ Stat.\ Phys.\ {\bf 71}, 967 (1993).

\bibitem{dirpoly} V.S.\ Dotsenko, L.B.\ Ioffe, V.B.\ Geshkenbein, 
                           S.E.\ Korshunov  and G.\ Blatter, 
    Phys.\ Rev.\ Lett. {\bf 100}, 050601 (2008)


\bibitem{Zhang} Yi-Cheng Zhang, 
   Europhys.\ Lett. {\bf 9}, 113 (1989)




\bibitem{KK1} I.V.\ Kolokolov and S.E.\ Korshunov, 
   Phys.\ Rev.\ B {\bf 75}, 140201(R) (2007);

\bibitem{KK2} I.V.\ Kolokolov and S.E.\ Korshunov, 
   Phys.\ Rev.\ B {\bf 78}, 024206 (2008); 

\bibitem{KK3} I.V.\ Kolokolov and S.E.\ Korshunov, 
   Phys.\ Rev.\ E {\bf 80}, 031107 (2009)



%%%%%%%%%%%%%%%%%%%%%%%%%%%%%%%%%%%%%%%%%%%%%%%%



\bibitem{Iwasaki} K.Iwasaki, H.Kimura, S.Shimomura and M.Yoshida,
   {\it From Gauss to Panleve\'e: a modern theory of special functions}.
   (Braunschweig, Vieweg 1991)

\bibitem{Hastings} S.P.Hastings and J.B.McLeod, 
   Arch.\, Rat.\, Mech.\, Anal.\, {\bf 73}, 31 (1980)













\end{thebibliography}
\end{document}